\documentclass[a4paper,12pt]{article}
\pdfoutput=1 

\usepackage{amsmath,amssymb}
\usepackage[pdftex,bookmarksnumbered=true]{hyperref}
\usepackage{color}
\hypersetup{colorlinks=true,linkcolor=blue,urlcolor=blue,filecolor=blue,citecolor=blue,pdfstartview=FitH,pdfpagemode=UseNone,bookmarksopen=true}
\usepackage[nocompress]{cite} 
\usepackage{fancyhdr}
\usepackage{enumerate} 

\usepackage{mathtools}
\usepackage[all]{hypcap}

\makeatletter

\newcommand{\av}{{\mathbf a}}

\newcommand{\xv}{{\mathbf x}}

\newcommand{\Fv}{{\mathbf F}}

\newcommand{\Lv}{{\mathbf L}}

\definecolor{cardinal}{rgb}{0.6,0,0}
\definecolor{darkgreen}{rgb}{0,0.4,0}
\definecolor{golden}{rgb}{0.92, 0.7, 0}
\definecolor{midnight}{rgb}{0, 0, 0.5}
\definecolor{darkblue}{rgb}{0, 0, 0.7}

\newcommand{\LG}[1]{\mathrm{#1}}

\DeclarePairedDelimiter\abs{\lvert}{\rvert}

\newcommand{\ud}[1]{\mspace{1mu}d#1\mspace{1mu}}
\newcommand{\vect}[1]{\mathbf{#1}}

\def\refchecklabelfontsize{\fontsize{5pt}{5pt}\selectfont}
\let\mark@size=\refchecklabelfontsize

\def\half{{\frac{1}{2}}}

\def\p{\partial}
\def\pb{\bar{\partial}}
\def\unit{{1\kern-.65ex {\rm l}}}
\def\1{{1\kern-.65ex {\rm l}}}
\def\ap{{\alpha'}}

\def\Im{\mathop{\mathrm{Im}}\nolimits}

\def\Re{\mathop{\mathrm{Re}}\nolimits}

\def\tr{\mathop{\mathrm{tr}}\nolimits}

\def\bracket#1{{\langle{#1}\rangle}}

\let\ev=\bracket

\def\zb{{\bar{z}}}

\def\Fb{{\bar{F}}}
\def\Gb{{\bar{G}}}
\def\Qb{{\bar{Q}}}

\def\cA{{\cal A}}

\def\cC{{\cal C}}

\def\cH{{\cal H}}

\def\cJ{{\cal J}}

\def\cN{{\cal N}}
\def\cO{{\cal O}}

\def\cQ{{\cal Q}}
\def\cR{{\cal R}}
\def\cS{{\cal S}}
\def\cT{{\cal T}}
\def\cU{{\cal U}}
\def\cV{{\cal V}}

\def\bbC{{\mathbb{C}}}

\def\bbR{{\mathbb{R}}}

\def\bbZ{{\mathbb{Z}}}

\newcount\hour \newcount\minute
\hour=\time \divide \hour by 60
\minute=\time
\def\now{%
\ifnum \hour<13
  \ifnum \hour=0 \advance \hour by 12 \number\hour:\else \number\hour:\fi%
     \ifnum \minute<10 0\fi%
     \number\minute%
\ A.M.%
\else \advance \hour by -12 \number\hour:%
  \ifnum \minute<10 0\fi%
  \number\minute%
  \ P.M.%
\fi%
}

\makeatother

\begin{document}

\baselineskip=18pt  
\numberwithin{equation}{section}  

\renewcommand{\headrulewidth}{0pt}
%
%


%



\thispagestyle{empty}

\vspace*{-2cm} 
\begin{flushright}
QMUL-PH-17-11\\
YITP-17-87
\end{flushright}

\vspace*{2.0cm} 
\begin{center}
 {\LARGE Non-Abelian Supertubes}\\
 \vspace*{1.7cm}
 Jos\'e J. Fern\'andez-Melgarejo$^{1,2}$,
 Minkyu Park$^1$ and Masaki Shigemori$^{1,3,4}$\\
 \vspace*{1.0cm} 
$^1$
Yukawa Institute for Theoretical Physics, Kyoto University,\\
Kitashirakawa-Oiwakecho, Sakyo-ku, Kyoto 606-8502 Japan\\[2ex]
$^2$
Departamento de F\'{\i}sica, Universidad de Murcia, E-30100 Murcia, Spain\\[2ex]
$^3$ Centre for Research in String Theory, School of Physics and Astronomy,\\
Queen Mary University of London, Mile End Road, London, E1 4NS, United Kingdom\\[2ex]
$^4$
Center for Gravitational Physics,
Yukawa Institute for Theoretical Physics, \\ Kyoto University,
Kitashirakawa-Oiwakecho, Sakyo-ku, Kyoto 606-8502 Japan\\
\medskip
\end{center}
\vspace*{1.0cm}

\noindent
%

\noindent
A supertube is a supersymmetric configuration in string theory which
occurs when a pair of branes spontaneously polarizes and generates a new
dipole charge extended along a closed curve.  The dipole charge of
a codimension-2 supertube is characterized by the U-duality monodromy as
one goes around the supertube.  For multiple codimension-2 supertubes,
their monodromies do not commute in general.
In this paper, we construct a supersymmetric solution of
five-dimensional supergravity that describes two supertubes with such
non-Abelian monodromies, in a certain perturbative expansion.
In supergravity, the monodromies are realized as the multi-valuedness of
the scalar fields, while in higher dimensions they correspond to
non-geometric duality twists of the internal space.
The supertubes in our solution carry NS5 and 5$_2^2$ dipole charges and exhibit
the same monodromy structure as the SU(2) Seiberg-Witten geometry.
The perturbative solution has ${\rm AdS}_2\times S^2$ asymptotics and
vanishing four-dimensional angular momentum. We argue that this
solution represents a microstate of four-dimensional black holes with a
finite horizon and that it provides a clue for the gravity realization
of a pure-Higgs branch state in the dual quiver quantum mechanics.

\newpage
\setcounter{page}{1} 



\tableofcontents


\section{Introduction and summary}

\subsection{Background}

The fact that black holes have thermodynamical entropy means that there
must be many underlying microstates that account for it.
Because string theory is a microscopic theory of gravity, i.e.,
quantum gravity, all these microstates must be describable within string
theory, at least as far as black holes that exist in string theory are
concerned.  A microstate must be a configuration in string theory with
the same mass, angular momentum and charge as the black hole it is
a microstate of, and the scattering  in the microstate must be
well-defined as a unitary process.
The fuzzball conjecture
\cite{Mathur:2005zp,Bena:2007kg,Skenderis:2008qn,Balasubramanian:2008da,Chowdhury:2010ct}
claims that typical microstates spread over a macroscopic distance of the
would-be horizon scale.  More recent arguments
\cite{Mathur:2009hf,Almheiri:2012rt} also support the view that the
conventional picture of black holes must be modified at the horizon
scale and replaced by some non-trivial structure.

The microstates for generic non-extremal black holes are expected to
involve stringy excitations and, to describe them properly, we probably
need quantum string field theory.  However, for supersymmetric black
holes, the situation seems much more tractable.  Many microstates for
BPS black holes have been explicitly constructed as regular, horizonless solutions of
supergravity --- the massless sector of superstring theory.  It is
reasonable that the massless sector plays an important role for
black-hole microstates because the large-distance structure expected of
the microstates can only be supported by massless fields
\cite{Gibbons:2013tqa}.
It is then natural to ask how many microstates of BPS black holes are
realized within supergravity.  This has led to the so-called
``microstate geometry program'' (see, e.g.,~\cite{Bena:2013dka}),
which is about explicitly constructing as many black-hole microstates as
possible, as regular, horizonless solutions in supergravity.

A useful setup in which many supergravity microstates have been
constructed is five-dimensional $\cN=1$ ungauged supergravity with
vector multiplets, for which all supersymmetric solutions have been
classified \cite{Gauntlett:2002nw, Bena:2004de}.  This theory describes
the low-energy physics of M-theory compactified on a Calabi-Yau 3-fold
$X$ or, in the presence of an additional $S^1$ \cite{Gauntlett:2004qy,
Gauntlett:2002nw}, of type IIA string theory compactified on $X$.  The
supersymmetric solutions are completely characterized by a set of
harmonic functions on a spatial $\bbR^3$ base, which we collectively
denote by $H$.  We will call these solutions \emph{harmonic solutions}.
If we assume that $H$ has codimension-3 singularities, its general form
is
\begin{equation}
 H(\xv)=h+\sum_{p=1}^N {\Gamma_p\over |\xv-\av_p|}.\label{Hharmonic}
\end{equation}
The associated supergravity solution generically represents a bound state of $N$
black-hole centers which sit at $\xv=\av_p$ ($p=1,\dots,N$) and are made
of D6, D4, D2, and D0-branes represented by the charge vectors
$\Gamma_p$.  In the current paper, we take $X=T^6=T^2_{45}\times
T^2_{67}\times T^2_{89}$ and the D-branes wrap some of the tori
directions.

By appropriately choosing the parameters in the harmonic functions, the
harmonic solutions with codimension-3 centers, \eqref{Hharmonic}, can
describe regular, horizonless 5D geometries that are microstates of black holes with
finite horizons \cite{Bena:2005va, Berglund:2005vb}.  However, although
they represent a large family of microstate geometries, it has been
argued that they are not sufficient for explaining the black-hole
entropy \cite{deBoer:2009un, Bena:2010gg}.

In fact, physical arguments naturally motivate us to generalize the
codimension-3 harmonic solutions, which leads to more microstates and
larger entropy.  One possible way of generalization is to go to six
dimensions.  This is based on the CFT analysis \cite{Bena:2014qxa} which
suggests that generic black-hole microstates must have traveling waves
in the sixth direction and thus depend on it. This intuition led to an
ansatz for 6D solutions \cite{Bena:2011dd}, based on which a new class
of microstate geometries with traveling waves, called superstrata, was
constructed~\cite{Bena:2015bea}. For recent developments in constructing
superstratum solutions,
see~\cite{Bena:2016agb,Bena:2016ypk,Tian:2016ucg,Bena:2017geu}.

The other natural way to generalize the codimension-3 harmonic solutions
\eqref{Hharmonic} is to consider \emph{codimension-2} sources in
harmonic functions.  This generalization is naturally motivated by the
supertube transition \cite{Mateos:2001qs} which in the context of
harmonic solutions implies that, when certain combinations of
codimension-3 branes are put together, they will spontaneously polarize
into a new codimension-2 brane.
For example, if we bring two orthogonal D2-branes together, they 
polarize into an NS5-brane along an arbitrary closed curve parametrized
by $\lambda$.  We represent this process by the following diagram:
\begin{align}
 \rm D2(45)+D2(67)\to ns5(\lambda 4567),
\label{D2D2ns5intro}
\end{align}
where D2(45) denotes the D2-brane wrapped on $T^2_{45}$ and ``ns5'' in
lowercase means that it is a dipole charge, being along a closed curve.
The original D2(45) and D2(67)-branes appeared in the harmonic functions
as codimension-3 singularities, as in \eqref{Hharmonic}.  The process
\eqref{D2D2ns5intro} means that those codimension-3 singularities can
transition into a codimension-2 singularity in the harmonic function
along the curve $\lambda$.
Another example of possible supertube transitions is
\begin{equation}
\rm D2(89)+D6(456789)\to 5^2_2(\lambda 4567;89),\label{D2D6522intro}
\end{equation}
where $5^2_2$ is a non-geometric exotic brane
\cite{Elitzur:1997zn,Blau:1997du,Hull:1997kb,Obers:1997kk,Obers:1998fb,deBoer:2010ud,deBoer:2012ma}
which is obtained by two transverse T-dualities of the NS5-brane
\cite{deBoer:2010ud, deBoer:2012ma}.

We emphasize that the supertube transition is not an option but a must;
if two codimension-3 branes that can undergo a supertube transition are
put together, they will, because the supertube is the intrinsic
description of the bound state \cite[Sec.~3.1]{Mathur:2005zp}.  This
suggests that considering only codimension-3 singularities in the
harmonic solutions is simply insufficient and we must include
codimension-2 supertubes for a full description of the physics.

In the presence of codimension-2 branes, the harmonic functions $H$ in
general become multi-valued \cite{Park:2015gka}.  This is because
codimension-2 branes generally have a non-trivial U-duality monodromy
around them \cite{deBoer:2010ud, deBoer:2012ma}, and $H$ transforms in a
non-trivial representation under it.  For a multi-center configuration,
if the $i$-th codimension-2 brane has U-duality monodromy represented by
a matrix $M_i$ around it, the harmonic functions will have the monodromy
\begin{align}
 H\to M_i H.
\end{align}
When the
matrices $M_i,M_j$ do not commute for some $i,j$, we say that the
configuration is \emph{non-Abelian}.\footnote{This is totally
different from making the gauge group non-Abelian, namely generalizing
Einstein-Maxwell to Einstein-Yang-Mills.  For some recent work on
non-Abelian generalizations in that sense,
see~\cite{Ramirez:2016tqc,Meessen:2017rwm}.}

In \cite{Park:2015gka}, two of the authors wrote down first examples of
codimension-2 harmonic solutions.  They involve multiple species of
codimension-2 supertubes and can have the same asymptotic charges as a
four-dimensional (4D) black hole with a finite horizon area.  However,
the constituent branes were \emph{unbound}; namely, by tuning parameters
of the solution, we can separate the constituents of the solution
infinitely far apart.  This implies that the solution does not actually
represent a microstate of a BPS black hole, for the following
reason~\cite[Sec.~3.1]{Mathur:2005zp}: Classically, it is possible to
consider a configuration in which constituents are separated by a finite
fixed distance from each other. However, quantum mechanically, by the
uncertainty principle, fixing the relative position of the constituents
increases kinetic energy and the configuration would not exactly
saturate the BPS bound. Namely, it cannot be a microstate of a BPS black
hole.  So, the solution constructed in~\cite{Park:2015gka} is not a
black-hole microstate.
Relatedly, the solution in~\cite{Park:2015gka} had Abelian monodromies.
There is some kind of linearity for codimension-2 branes with commuting
monodromies, and we can construct solutions with multiple codimension-2
centers basically by adding harmonic functions for each
center.\footnote{More precisely, one should include certain interaction
terms as well \cite{Park:2015gka}. However, it is still true in this
case that one can in principle construct solutions with multiple
codimension-2 centers located wherever we want. \label{ftnt:linearity}} This
suggests that codimension-2 branes with Abelian monodromies do not talk
to each other and are not bound.

Then the natural question is: does a configuration of supertubes with
non-Abelian monodromies exist?  If so, is it a bound state, and does it
represent a black-hole microstate?  These are precisely the questions
that we address in this paper.

\subsection{Main results}

In this paper, we will construct a configuration of codimension-2
supertubes with \emph{non-Abelian monodromies} within the framework of
harmonic solutions, in a certain perturbative expansion.  We will give
evidence that, as expected, it represents a bound state, and that it
corresponds to a microstate of a 4D black hole with a finite horizon.

Our configuration is made of two circular supertubes which share their
axis.  The two tubes are separated by distance $2|L|$ and the radii of
both rings are approximately $R$.
See Figure~\ref{fig:colliding} on page  \pageref{fig:colliding}.
The harmonic functions $H$ will have a non-trivial monodromy around each
of the two tubes.  The monodromies for the two supertubes do not
commute, namely, they are non-Abelian.
Because it is technically difficult to find the solution for general $R$
and $|L|$, we consider the ``colliding limit'', $|L|\ll R$, in which we
can construct the harmonic functions order by order in a perturbative
expansion.

Despite that the colliding limit allows us to construct the solution
explicitly, it also has a drawback: we cannot determine the value of $R$
and $|L|$ separately.  If we knew the exact solution, not a perturbative
one, then we would be able to constrain them by imposing physical conditions
(the absence of closed timelike curves) on the explicit solution. In
this paper we will not be able to do that.  Instead, we will make use of
supertube physics to argue that $R$ and $|L|$ are fixed (Section
\ref{ss:arg_bnd_state}).  Although the argument physically well
motivated and convincing, it is not a proof; we hope to revisit this
point in future work.

Because the physical parameters $R$ and $|L|$ are fixed, it
is not possible to separate apart the two supertubes and therefore the
configuration represents a bound state.  Moreover, it has asymptotic
charges of a 4D black hole with a finite horizon.  Therefore, the
non-Abelian 2-supertube configuration is arguably a black-hole
microstate.  The geometry is not regular near the supertubes, but the
singular behavior is an allowed one in string theory, just as the
geometry near a 1/2-BPS brane is metrically singular but is allowed.  In
this sense, our solution is not a microstate geometry but a microstate
solution as defined in \cite{Bena:2013dka}.  Our solution simultaneously
involves the two types of supertube, \eqref{D2D2ns5intro} and
\eqref{D2D6522intro}, and therefore is non-geometric in that the
internal torus is twisted by T-duality transformations around the
supertubes.

We find that the asymptotic geometry of the perturbative solution is
AdS$_2\times S^2$, namely the attractor geometry \cite{Ferrara:1995ih}
of the black hole with the same charge. Furthermore, we find that the 4D
angular momentum of the solution is zero, $J=0$.  We will argue that
this is due to cancellation between the angular momentum that the
individual supertubes carry and the one coming from the electromagnetic
crossing between the monopole charges carried by the supertubes.

On a more technical note, in the colliding limit $|L|\ll R$, we can
split the problem of finding harmonic functions with desired monodromies
into two parts.  If one is at a distance $d\sim R\gg |L|$ away from the
supertubes (the ``far region''), the configuration is effectively
considered as made of a single tube whose monodromy is the product of
two individual monodromies.  On the other hand, if one is at a distance
$d\sim |L|\ll R$ away from the tubes (the ``near region''), we can
regard the tubes as infinitely long and the problem reduces to that of
finding 2D harmonic functions with desired monodromies.  Once we find
harmonic functions in both regions, we can match them order by order in
a perturbative expansion to construct the harmonic function in the
entire space.  This is the sense in which our solution is perturbative
in nature.
In the near region, the problem is to find a pair of holomorphic functions with non-trivial
$\LG{SL}(2,\bbZ)$ monodromies around two singular points on the complex
$z$-plane. Mathematically, this problem is the same as the one
encountered in the $\LG{SU}(2)$ Seiberg-Witten theory
\cite{Seiberg:1994rs} and we borrow their results to construct the
harmonic functions.

The solution thus constructed is perfectly consistent at the
perturbative level, but it is possible that unexpected new features are
encountered in the exact, full-order solution.  However, constructing
such an exact solution is beyond the techniques developed in this paper
and left for future research.


In terms of the harmonic solutions $H=\{V,K^I,L_I,M\}$, our
configuration is given by
\begin{align}
\begin{aligned}
V &= \Re G \, ,&\quad  K^1 &=K^2=-\Im G, &\quad  K^3&= \Re F\, ,\quad &&
\\
&&L_1&=L_2=\Im F\,,  & L_3&=\Re G\, ,& M&=-\frac12 \Re F
\, ,
\end{aligned}
\end{align}
where $F$ and $G$ are complex functions and carry the information of the
monodromies. This class of solutions describes the general configuration
in which the complexified K\"ahler moduli of $T^2_{45}$ and $T^2_{67}$
are set to $\tau^{1,2}=i$ whereas the one associated with $T^2_{89}$ is
given by $\tau^3=\frac FG$.  This class is a type IIA realization of the
so-called SWIP solution  \cite{Bergshoeff:1996gg}.
It is the particular choice of
the pair $(\begin{smallmatrix}F\\ G\end{smallmatrix})$ that fixes the
monodromies of the configuration. In our solution, $F$ and $G$ are
related to the defining functions of the Seiberg-Witten solution.


\subsection{Implication for black-hole microstates}

In the above, we argued that our codimension-2 configuration represents
a black-hole microstate.  Our perturbative solution is quite different
from the supergravity microstates based on co\-di\-men\-sion-3 harmonic
solutions \cite{Bena:2005va,Berglund:2005vb,Bena:2007kg} that have been
extensively studied in the literature.
In particular, its 4D asymptotics is the
AdS$_2\times S^2$ attractor geometry of the black hole with the same
asymptotic charges, because the harmonic functions cannot have constant
terms.  Furthermore, the 4D angular momentum of our solution
vanishes, $J=0$, because of a cancellation
mechanism between the tube and crossing contributions.
To better understand the possible implications of these properties, let
us recall some known facts and conjectures about black-hole microstates.

For codimension-3 harmonic solutions, a well-known family of microstate
geometries whose 4D asymptotics can be made arbitrarily close to
AdS$_2\times S^2$ and whose angular 4D momentum $J$ can be made
arbitrarily small is the so-called scaling solutions
\cite{Denef:2002ru,Bena:2006kb,Bena:2007qc}.\footnote{Note that the
angular momentum here is the 4D one.  In the scaling solution, the 4D
angular momentum can be made arbitrarily small.  If one goes to 5D,
there are two angular momenta, and the 4D angular momentum is one of the
two.  The other 5D angular momentum, which is nothing but the D0-brane
charge from the 4D viewpoint, has been quite difficult to make smaller
than a certain lower limit, for the geometry to correspond to a
microstate in the D1-D5 system
\cite{Bena:2006kb,Bena:2007qc,Heidmann:2017cxt,Bena:2017fvm}. This problem can be
overcome by generalizing the harmonic solution to the superstratum in 6D
\cite{Bena:2016ypk}.  This issue is not relevant to the current
discussion.} Scaling solutions are made of three or more codimension-3
centers and exist for any value of the asymptotic moduli, provided that
certain triangle inequalities are satisfied by the skew products of the
charges of the centers.  The defining property of the scaling solutions
is that we can scale down the distance between centers in the $\bbR^3$
base so that they appear to collide.  However, the actual geometry does
not collapse; what is happening in this scaling process is that an AdS
throat gets deeper and deeper, at the bottom of which the non-trivial
2-cycles represented by the centers sit.  At the same time, the angular
momentum $J$ becomes smaller and smaller.  In the infinite scaling limit
where all the centers collide in the $\bbR^3$ base, the geometry becomes
precisely AdS and the angular momentum $J$ vanishes.
It has been argued~\cite{Martinec:2015pfa,Martinec:2017ztd} that the
majority of the black-hole microstates live in this infinite scaling
limit, where the branes wrapping the 2-cycles~\cite{Denef:2007yt},
called ``W-branes'', become massless and condense.  In the IIA picture,
W-branes are fundamental strings stretching between D-brane centers. In
the language of quiver quantum mechanics~\cite{Denef:2002ru} dual to
scaling solutions, the configurations with a finite throat correspond to
Coulomb branch states, while the configurations with W-brane condensate
would correspond to pure-Higgs branch states~\cite{Bena:2012hf}.
However, the gravity description of such condensate is
unclear.\footnote{For recent attempts to construct the gravity
description of W-branes, see
\cite{Levi:2009az,Raeymaekers:2014bqa,Raeymaekers:2015sba,Tyukov:2016cbz}.
}  It cannot simply be the infinite throat limit of the scaling
solution, because in that limit the non-trivial 2-cycles disappear in the
infinite depth and the entire geometry becomes just AdS,
indistinguishable from the black-hole geometry.
Furthermore, quantization of the solution space of the scaling solutions
\cite{deBoer:2008zn} says that the depth of the throat cannot be made
arbitrarily large but is limited by quantum effects.
So, it appears that, although the scaling solution is an important clue
for the W-brane condensate and pure-Higgs branch states, it is not the
answer itself.

Relatedly, Sen and his collaborators argued
\cite{Sen:2009vz,Dabholkar:2010rm,Chowdhury:2015gbk} that the
contribution to black-hole microstates can be split into the ``hair''
part which lives away from the horizon and the ``horizon'' part which
gives the main contribution to black-hole entropy.  The horizon part has
asymptotically AdS$_2$ geometry and vanishing angular momentum, $J=0$.
This is based on the fact that, in 4D, only $J = 0$ black holes are BPS
and all extremal black holes with $J\neq 0$ are
non-supersymmetric~\cite{Sen:2009vz}. The analysis of the quiver quantum
mechanics describing the worldvolume theory of a D-brane black-hole
system~\cite{Chowdhury:2015gbk} also supports the claim that all
black-hole microstates in 4D have $J=0$.

In summary, both the analysis of the scaling solutions and the arguments of Sen et al.
suggest that the majority of the black-hole microstates have
AdS asymptotics and vanishing angular momentum, $J=0$.  They are states
with a condensate of W-branes, or equivalently fundamental strings
stretching between D-branes, and correspond to the pure-Higgs branch
states of the dual quiver quantum mechanics.

Now if we look at our perturbative solution, it seems to have all the
above properties expected of a typical microstate of a 4D black hole.
First, it has AdS$_2$ asymptotics. This was not done by
fine-tuning of parameters but is a consequence of the non-trivial
monodromy of the supertubes.  Second, its angular momentum vanishes,
$J=0$.  This did not require  fine-tuning either, and it was due to
the cancellation mechanism mentioned before between different
contributions to angular momentum.
Moreover, our solutions are made of supertubes generated by the
supertube transition which is nothing but condensation of the strings
stretching between the original D-branes.  Therefore, it is natural to
conjecture that our solution is giving a gravity description of the
W-brane condensate and represents a state in the pure-Higgs branch.  At
least, it is expected to provide a clue for the gravity description of
pure-Higgs branch states.

Of course, to make such a strong claim we need strong evidence,
including the demonstration that non-Abelian supertube configurations do
exist beyond the perturbative level, and the proof they have a huge
entropy to account for the black-hole microstates.  Such studies would
require more sophisticated tools and techniques than developed in the
current paper.  At this point, we just state that it is quite
non-trivial and intriguing that the perturbative non-Abelian 2-supertube
solution has the properties expected of black-hole microstates, and
leave further investigation as an extremely interesting direction of
future research.

In \cite{Lunin:2015hma} (see also \cite{Pieri:2016cqz}), an interesting
set of solutions with ${\rm AdS_2}\times S^2$ asymptotics were
constructed.  They belong to the so-called IWP family of solutions
\cite{Israel:1972vx, Perjes:1971gv} and are characterized by one complex
harmonic function in three dimensions.  The main differences between the
solutions in \cite{Lunin:2015hma} and ours are as follows.  First,
because the solutions in \cite{Lunin:2015hma} are based on one complex
harmonic function, their possible monodromies are Abelian.  On the other
hand, our solution has two complex harmonic functions and thus the
monodromies are in general non-Abelian.  Second, the solutions in
\cite{Lunin:2015hma} have two distinct ${\rm AdS_2}\times S^2$
asymptotic regions. In contrast, the multiple asymptotic regions in our
solutions are related by U-duality and regarded as one asymptotic
region in different U-duality frames.  Therefore, our solution has
only one physical asymptotic region.

Let us end this section by mentioning one other difference between
microstates with codimension-3 centers and ones with codimension-2
centers.
One issue about the existing construction of black-hole microstates
based on codimension-3 harmonic solutions is that, multi-center
configurations (except for the case where there are two centers and one
of them is a 1/2-BPS center) are expected to lift and disappear from the
BPS spectrum once generic moduli are turned on \cite{Dabholkar:2009dq}.
The physical origin of this is that, if there are multiple centers, when
one continuously changes the moduli to arbitrary values, the discreteness
of quantized charges is incompatible with the BPS condition
\cite{Chowdhury:2013pqa}.  This is certainly an issue for codimension-3
centers but, codimension-2 supertubes may be able to avoid it by
continuously deforming the tube shapes and re-distributing the monopole
charge density along its worldvolume, so that the BPS condition is met
even if one changes the moduli continuously.  Therefore, it may be that
codimension-2 solutions provide a loophole for the no-go result of
\cite{Dabholkar:2009dq} and represent microstates that remain
supersymmetric everywhere in the moduli space.

\subsection{Plan of the paper}

The rest of this paper is organized as follows.
In Section \ref{sec:harmonic} we review the BPS solutions, called
harmonic solutions, which can describe a wide class of multi-center
configurations in string theory in four and five dimensions.  We discuss
their physical properties, giving examples for cases with codimension-3
and codimension-2 centers.  We also introduce the class of solutions in
which only one $\text{SL}(2,\bbZ)$ duality is turned on and has only one
modulus $\tau$.
In Section \ref{sec:examnonAbel}, we explicitly construct an example of
non-Abelian supertubes.  We first introduce the colliding limit and the
matching expansion which allow us to construct the solution order by
order by connecting the far-region and near-region solutions.  We then
use it to perturbatively construct the solution.  As the near-region
solution, we use an ansatz inspired by the $\LG{SU}(2)$ Seiberg-Witten
theory.
In Section \ref{sec:physprop}, we study the physical properties of the
solution.  We work out the brane charge content, the asymptotic geometry
and the angular momentum, and discuss the condition for the absence of
closed timelike curves (CTCs).  Based on the results, we argue that the
solution is a bound state and thus represent a black-hole microstate.
We also discuss the cancellation mechanism responsible for the
vanishing  of the angular momentum.

The Appendices include some details of the computations carried out in the main
text and some topics tangential to the content of the main text.
In Appendix \ref{app:duality_H}, we discuss some aspects of the duality
transformations acting on the harmonic functions.
In Appendix \ref{app:perturb}, we discuss some details of the matching
expansion to higher order than is discussed in the main text.
In the main text, we focus on the class of solutions in which only one
of the three moduli of the STU model is activated.  In Appendix
\ref{app:tau1=i}, we discuss the class of solutions in which two of
moduli are activated.
In Appendix \ref{app:supertubes}, we discuss properties of the
supertubes created from a general 1/4-BPS center in the one-modulus
class of solutions.
In Appendix \ref{app:D2+D6->522}, we present the explicit harmonic
functions for the $\text{D2}+\text{D6}\to 5^2_2$ supertube used in the
main text.

\section{Multi-center solutions with codimension 2 and 3}
\label{sec:harmonic}

\subsection{The harmonic solution}
\label{sec:uplift-dualities}

The most general supersymmetric solutions of ungauged $d=5$, $\cN=1$
supergravity with vector multiplets have been classified in
\cite{Gutowski:2004yv} (see also \cite{Gauntlett:2002nw, Bena:2004de,
Gutowski:2005id}).\footnote{Depending on whether the Killing vector
constructed from the Killing spinor bilinear is timelike or null, the
solutions are classified into timelike and null classes.  In this paper
we will consider the timelike class.} When one applies this result to
M-theory compactified on $T^6=T^2_{45}\times T^2_{67}\times T^2_{89}$
(the so-called STU model) and further assumes a tri-holomorphic
$\LG{U}(1)$ symmetry \cite{Gauntlett:2004qy}, the general supersymmetric
solution corresponds to the following 11-dimensional fields:
\begin{align} 
\label{Msol}
\begin{aligned}
ds_{11}^2
&=
-Z^{-2/3}(dt+k)^2
+Z^{1/3}ds_{\mathrm{GH}}^2
+Z^{1/3}\left(Z_1^{-1}dx_{45}^2
+Z_2^{-1}dx_{67}^2+Z_3^{-1}dx_{89}^2\right) 
\, ,
\\[.5ex]
\cA_3
&=
\left(B^I - Z_I^{-1}(dt+k)\right)\wedge  J_I
\, ,
\quad
J_1 \equiv dx^4\wedge dx^5\,,~
J_2 \equiv dx^6\wedge dx^7\,,~
J_3 \equiv dx^8\wedge dx^9\,,
\end{aligned}
\end{align}
where $I=1,2,3$; $Z\equiv Z_1 Z_2 Z_3$; and $dx_{45}^2\equiv
(dx^4)^2+(dx^5)^2$ etc.

Supersymmetry requires that all fields in \eqref{Msol} be written in
terms of \emph{3D harmonic functions} as follows
\cite{Gauntlett:2004qy}.  First, the metric $ds_{\mathrm{GH}}^2$ must be
a 4-dimensional metric of a Gibbons-Hawking space given by
\begin{align}
ds_{\mathrm{GH}}^2
=
V^{-1}(d\psi+A)^2
+V d\xv^2
\, , \qquad 
 \psi\cong \psi+4\pi\,,\qquad
\xv=(x^1,x^2,x^3)
\, .
\label{GHmetric}
\end{align}
The 1-form $A$ and the scalar $V$ depend on
the coordinates $\xv$ of the $\mathbb{R}^3$ base and satisfy
\begin{align}
dA= *_3\, dV
\, ,
\end{align}
where $*_3$ is the Hodge dual operator on the $\mathbb{R}^3$. From this, we see that $V$ has to be a harmonic function in
$\mathbb{R}^3$,
\begin{align}
\Delta V 
= 
0 
\, ,
\qquad\qquad
\Delta 
\equiv
\partial_i\partial_i
\, .
\end{align}
The rest of the fields can be written in terms of additional harmonic
functions $K^I,L_I,M$ on $\bbR^3$ as follows:
\begin{align}
B^I &= V^{-1} K^I (d\psi+ A)+ \xi^I
\, ,
\qquad\qquad
d \xi^I
=
- *_3 dK^I
\, ,\\
Z_I 
&=
L_I+\frac12 C_{IJK} V^{-1} K^J K^K
\, ,
\\
k 
&=
\mu (d\psi + A) 
+\omega
\, ,
\label{k1form}
\\
\mu 
&=M+
\frac12 V^{-1} K^I L_I
 +\frac16 C_{IJK} V^{-2} K^I K^J K^K
\, ,
\end{align}
where $C_{IJK}=|\epsilon_{IJK}|$. If one replaces the internal space
$T^6=(T^2)^3$ by a Calabi-Yau 3-fold $X$, most of our formulas remain
valid as long as we replace $C_{IJK}$ by the triple intersection numbers
of $X$ \cite{Gauntlett:2004qy}.
We sometimes write eight harmonic functions collectively as
$H=\{V,K^I,L_I,M\}$.  For two such vectors
$H,H'$, we define the skew product $\bracket{H,H'}$ by
\begin{align}
 \bracket{H,H'}\equiv VM'-MV'+{1\over 2}(K^I L_I'-L_I K'^I).\label{def_<H,H>}
\end{align}

The 1-form $\omega$ satisfies
\begin{align}
 *_3 d \omega 
 &= \bracket{H,dH}
 \, .
 \label{sdomega}
\end{align}
Applying $d\,*_3$ on this equation implies
\begin{align}
0 
 &= \bracket{H,\Delta H}
\, .
\label{integrability}
\end{align}
This is often called the \emph{integrability condition}
\cite{Denef:2000nb} (see also \cite{Bena:2005va}), and is a
necessary requirement for the existence of $\omega$. Harmonicity of the
functions $H=\{ V,K^I, L_I, M\}$ may make one think that
the right-hand side identically vanishes. However, the harmonic
functions generically have singularities associated with the presence of
sources, which can lead to a non-vanishing contribution to the right-hand
side and make $\omega$ multi-valued.  Whether we must allow or disallow
such contribution must be determined based on physical considerations,
as we will discuss below in concrete examples.

The above represent a  broad family of supersymmetric solutions
characterized by eight harmonic functions, $H=\{ V,K^I,L_I,M \}$. We call this set of solutions \emph{harmonic
solutions}.\footnote{These solutions were first found in
\cite{Behrndt:1997ny} as solutions of $d=4,\cN=2$ supergravity with
vector multiplets and made more explicit in \cite{Bates:2003vx}. In 5D,
the supersymmetric solutions in $\cN=2$ supergravity with vector
multiplets in the timelike class were classified in
\cite{Gutowski:2004yv, Bena:2004de} (see also \cite{Gauntlett:2002nw,
Gutowski:2005id}) and later reduced to 4D solutions in
\cite{Gauntlett:2004qy}, which are identical to the ones in
\cite{Behrndt:1997ny,Bates:2003vx}.  In 4D, it was later shown in
\cite{Meessen:2006tu} that these solutions are the most general
supersymmetric solutions in the timelike class in $d=4,\cN=2$ ungauged
supergravity with vector multiplets.  There being no widely accepted
name for these solutions, we call them harmonic solutions.}
Although we started with $d=5$ supergravity, the existence of the
isometry along $\psi$ allows us to dimensionally reduce the solution to
4D\@.  Therefore, the harmonic solutions can be regarded as representing
configurations in 4D\@.


\bigskip
Reducing the 11D solution \eqref{Msol} along $\psi$, we obtain the
following supersymmetric solution of type IIA
supergravity:\footnote{For expressions for higher RR potentials, see,
e.g.,~\cite[App.~E]{Park:2015gka} and \cite{DallAgata:2010srl}.}
\begin{align}
ds_{10,\mathrm{str}}^2
&=
-\frac{1}{\sqrt{\cQ\,}}(dt+\omega)^2
+\sqrt{\cQ\,} \, d\xv^2
+ \frac{\sqrt{\cQ\,}}{V}\left(Z_1^{-1}dx_{45}^2
+Z_2^{-1}dx_{67}^2+Z_3^{-1}dx_{89}^2\right)
\, ,
\notag
\\[.55em]
e^{2\Phi}
&=
\frac{\cQ^{3/2}}{V^{3} Z}
\, ,\qquad
B_2
=
\left(
	V^{-1} K^I
	-Z_I^{-1}  \mu
	\right) J_I\, ,
\label{IIAfield}
\\[.55em]
C_1
&=
A
-\frac{V^2\mu}{\cQ}(dt+\omega)
\, ,\qquad
C_3
=
\left[
	(V^{-1}K^I-Z_I^{-1}\mu) A
	+\xi^I-Z_I^{-1}(dt+\omega)
	\right] \wedge J_I 
\, ,
\notag
\end{align}
where $ds^2_{\rm 10,str}$ is the string-frame metric and
\begin{align}
\cQ \equiv V(Z-\mu^2 V)
\, .
\label{Q_def}
\end{align}
Explicitly in terms of harmonic functions,
\begin{align}
 \cQ&=
  V L_1 L_2 L_3 - 2 M K^1 K^2 K^3
 - M^2 V^2\notag\\
 &\qquad
 -{1\over 4} \sum_{I}(K^I L_I)^2
 + {1\over 2} \sum_{I<J} K^I L_I K^J L_J
 -  M V \sum_I K^I L_I
 \notag\\
 &\equiv J_4(H)
 ,
\label{J4_def0}
\end{align}
where $J_4$ is the quartic invariant of the STU model; for some more
discussion, see Appendix~\ref{app:duality_H}\@.

Let the complexified K\"ahler moduli for the 2-tori $T^2_{45}$,
$T^2_{67}$, and $T^2_{89}$ be $\tau^1$, $\tau^2$, and $\tau^3$,
respectively.  The expression in terms of harmonic functions is
\begin{align} 
\tau^1
 =B_{45}+i\sqrt{\det G_{ab}}
 =\left({K^1\over V}-{\mu\over Z_1}\right)+i\frac{\sqrt{\cQ\,}}{Z_1V}\,,
 \label{kahlerm}
\end{align}
where $a,b=4,5$ and the radii of 456789 directions have been all set to
$l_s=\sqrt{\alpha'}$.  The other moduli $\tau^2$ and $\tau^3$ are given
by the same expression with $45$ replaced by $67$ and $89$,
respectively.  In supergravity, these moduli parametrize the moduli
space $\big[{\LG{SL}(2,\bbR)\over\LG{SO}(2)}\big]^3$.  In string theory, this
reduces to $\big[{\LG{SL}(2,\bbR)\over\LG{SL}(2,\bbZ)\times\LG{SO}(2)}\big]^3$ by
the $[\LG{SL}(2,\bbZ)]^3$ duality symmetry that identifies different values
of $\tau^I$.

%

For other embeddings of the harmonic solutions
in type IIA and IIB supergravity, see~\cite{Bena:2004de, Elvang:2004ds, Bena:2008dw}.

\subsubsection*{Duality transformations}

Because we will consider codimension-2 configurations with non-trivial
U-duality monodromies, it is useful to recall some facts about the
U-duality group in the STU model, which is $\text{SL}(2,\bbZ)_1\times
\text{SL}(2,\bbZ)_2\times \text{SL}(2,\bbZ)_3$ \cite{Duff:1995sm}.

In particular, it is important to understand how the U-duality acts on
the harmonic functions. Let us take $\text{SL}(2,\bbZ)_1$. This
group is generated by (i) simultaneous T-duality transformations on the
45 directions and (ii) the shift symmetry $B_{45}\to B_{45}+1$.  Because we know the T-duality action on 10D fields
from the Buscher rule and their expression \eqref{IIAfield} in terms of
harmonic functions, it is easy to read off how the harmonic functions
transform under (i). The same is true for the $B$-shift symmetry~(ii).
The result is that (i) and (ii) are realized by the $\text{SL}(2,\bbZ)_1$
matrices
\begin{align}
 M_{\text{T-duality}}=\begin{pmatrix} 0 &-1\\ 1 & 0  \end{pmatrix},\qquad
 M_{\text{$B$-shift}}=\begin{pmatrix} 1&1\\0&1   \end{pmatrix}\,,
\end{align}
and that the eight harmonic functions transform as a direct sum of four
doublets,
\begin{align}
\begin{pmatrix}K^1 \\ V\end{pmatrix}\, ,\quad
\begin{pmatrix}2M \\ -L_1\end{pmatrix}\, ,\quad
\begin{pmatrix}-L_2 \\ K^3\end{pmatrix}\, ,\quad
\begin{pmatrix}-L_3 \\ K^2\end{pmatrix}\, .
\label{SL21doublets}
\end{align}
Since (i) and (ii) generate $\text{SL}(2,\bbZ)_1$, we conclude that, even for
general transformations $\text{SL}(2,\bbZ)_1$, the harmonic functions transform
as a collection of doublets \eqref{SL21doublets}.

Because all three $\text{SL}(2,\bbZ)$'s are on the same footing, we can
infer the transformation of harmonic functions under general
$\text{SL}(2,\mathbb{R})_I$ transformation for $I=1,2,3$.   Under
$\text{SL}(2,\mathbb{R})_I$, the eight harmonic functions transform as a
direct sum of four doublets:
\begin{align}
\begin{pmatrix}
u \\ v
\end{pmatrix}
\to
M_I
\begin{pmatrix}
u \\ v
\end{pmatrix}
\, ,
\qquad
M_I\equiv
\begin{pmatrix}
\alpha_I & \beta_I
\\
\gamma_I & \delta_I
\end{pmatrix}
\in
\text{SL}(2,\mathbb{R})_I,
 \label{SL2transf_harmfunc}
\end{align}
where the vector $(\begin{smallmatrix}u\\ v
\end{smallmatrix})$ represents any of the pairs
\begin{align}
\begin{pmatrix}K^I \\ V\end{pmatrix}\, ,\quad
\begin{pmatrix}2M \\ -L_I\end{pmatrix}\, ,\quad
\begin{pmatrix}-L_J \\ K^K\end{pmatrix}\, ,\quad
\begin{pmatrix}-L_K \\ K^J\end{pmatrix}\, ,
\qquad
J\neq K \neq I
\, .
\label{SL2doublets}
\end{align}
One can show that the transformations \eqref{SL2transf_harmfunc} for
different values of $I$ commute, as they should because they are
associated with different tori.

It is not difficult to show that the transformation
\eqref{SL2transf_harmfunc} of the harmonic functions means the standard
linear fractional transformation of the  complexified K\"ahler moduli as:
\begin{align}
\tau^I 
\to&\
\frac{\alpha_I\, \tau^I+\beta_I }{\gamma_I \,\tau^I + \delta_I}
\, ,
\qquad
\tau^J\to \tau^J \quad ( J\neq I)
\, ,
\label{tau_transf}
\end{align}
where there is no summation over $I$.
%

For some more aspects on the duality transformation of the harmonic
solutions, see Appendix \ref{app:duality_H}\@.


\subsubsection*{Conditions for the absence of closed timelike curves}

(Super)gravity solutions can exhibit closed timelike curves (CTCs),
signaling that the solution is not physically allowed.\footnote{For over-rotating supertubes, CTCs can appear along the profile of the supertube \cite{Emparan:2001ux, Mateos:2001pi}.}
To study their existence, let us
look at the 10D metric \eqref{IIAfield}.
First, for $g_{tt},g_{ii}$ $(i=1,2,3)$ to be real, we need $\cQ\ge 0$.
Then, for the torus directions to give no CTCs, we get $V Z_I\ge0$,
$I=1,2,3$.  So, we must impose the following conditions:
\begin{subequations}\label{condCTCs}
\begin{align}
\cQ 
&\ge
0
\, ,
\label{QcondCTCs}
\\
V Z_I
&\ge
0
\, .
\label{VZcondCTCs}
\end{align}
\end{subequations}

Next, let us focus on the $\bbR^3$ part of the 10D metric
\eqref{IIAfield} which is
\begin{align}
ds_{10,\text{str}}^2\supset-{\omega^2\over \sqrt{\cQ}}+\sqrt{\cQ}\,d{\bf x}^2
 .
 \label{omegacondCTCs}
\end{align}
It is possible that closed curve $\cC$ in $\bbR^3$ becomes timelike
under this metric, depending on the behavior of the 1-form
$\omega$. That would imply a CTC, which must be physically disallowed.  We
will discuss this condition in specific situations later.



\subsection{Configurations with only one modulus}
\label{sec:taucases}

Thus far, we have been discussing configurations for which all moduli
$\tau^I$, $I=1,2,3$ can in principle be all non-trivial.  Now let us
focus on configurations with
\begin{align}
 \tau^1=\tau^2=i,\qquad \tau^3=\text{arbitrary}.
\label{tau1=tau2=i}
\end{align}
Although being particular instances of the general solution, they can
still describe a wide range of physical configurations, such as ones
with multiple centers with codimension 3 and 2. This class of solutions
provides a particularly nice setup for our purpose of constructing
codimension-2 solutions with non-Abelian monodromies.  This class is
nothing but a type IIA realization of the solution called the SWIP
solution in the literature \cite{Bergshoeff:1996gg}.  Here we discuss
some generalities about this class.

Using the expression \eqref{kahlerm} for $\tau^I$ in terms of harmonic
functions, we see that the condition \eqref{tau1=tau2=i} implies the following relations:
\begin{align}
K^1=K^2\,,\qquad
L_1=L_2\,,\qquad
L_3=V\,,\qquad
M=-\frac{K^3}{2}\,,\label{tau12iharmcond}
\end{align}
leaving four independent harmonic functions. If we plug these expressions
into \eqref{kahlerm}, we obtain
\begin{align}
\tau^3
=
\frac{K^3+i L_1}{V-i K^1}={F\over G}
\, ,\label{tau3=F/G}
\end{align}
%
where we defined complex combinations
\begin{align}
F\equiv K^3+i L_1,\qquad
G\equiv V-iK^1.
\label{FG_def}
\end{align}
As we can see from \eqref{SL2doublets}, the pair
$(\begin{smallmatrix}F\\ G\end{smallmatrix})$ transforms as a (complex) doublet
under $\text{SL}(2,\bbZ)_3$. From the expression \eqref{tau3=F/G}, it is obvious that $\tau^3$ undergoes
linear fractional transformation under $\text{SL}(2,\bbZ)_3$ (although
we already said this in \eqref{tau_transf} in general).
The harmonic
functions are written in terms of them as
\begin{align}
\begin{aligned}
V &= \Re G \, ,&\quad  K^1 &=K^2=-\Im G, &\quad  K^3&= \Re F\, ,\quad &&
\\
&&L_1&=L_2=\Im F\,,  & L_3&=\Re G\, ,& M&=-\frac12 \Re F
\, .
\end{aligned}
\label{tau12harm}
\end{align}
%
In terms of the complex quantities $F,G$, some previous formulas become
\begin{align}
 \bracket{H,H'}&=\Re(F\Gb'-G \Fb'),\label{skew_prod_FG}\\
 \cQ&=(\Im F\bar{G})^2.\label{Q_FG}
\end{align}
The equation for $\omega$, \eqref{sdomega}, reads
\begin{align}
*_3d\omega 
=
\Re \left(
	F d \bar{G}
	- G d \bar{F}
	\right)
\, .
\label{domega_FG}
\end{align}

Let us consider the general no-CTC conditions.  Under the constraint
\eqref{tau12iharmcond}, the condition \eqref{QcondCTCs} is automatically
satisfied because $\cQ=(\Im F\bar{G})^2\ge 0$. On the other hand, the
condition \eqref{VZcondCTCs} gives
\begin{align}
\Im(F \bar{G}) 
 =|G|^2\Im \tau^3
 \ge 0
\, .
\label{tau12CTCs}
\end{align}
%


Here we have seen that switching off two moduli $\tau^1$ and $\tau^2$ leads to
a substantial simplification.  In Appendix~\ref{app:tau1=i}, we discuss
switching off one modulus $\tau^1$, which also leads to interesting
simplification.

\subsection{Codimension-3 solutions}
\label{sec:codim3}

The harmonic solutions are characterized by a set of 8 harmonic
functions.  Non-trivial harmonic functions in $\bbR^3$ must have
singularities, which correspond to physical sources such as D-branes.
Depending on the nature of the source, the singularity can have various
codimension.  Here we review some specifics about solutions with
codimension-3 sources, or codimension-3 solutions for short, which have
been extensively studied in the literature.  In the next subsection, we
will proceed to codimension-2 solutions, which is the main focus of the
current paper.

If one assumes that all singularities of the harmonic functions have
codimension 3, the general form of the harmonic functions is
\cite{Behrndt:1997ny, Bates:2003vx, Gauntlett:2004qy}
\begin{align}
\begin{aligned}
V  
&=
h^0+\sum_{p=1}^N {\Gamma^0_p\over |\xv-\av_p|}
\,,
&\qquad
K^I
&=
h^I+\sum_{p=1}^N {\Gamma^I_p\over |\xv-\av_p|}
\,,
\\
L_I
&=
h_I+\sum_{p=1}^N {\Gamma_I^p\over |\xv-\av_p|}
\,,
&
M
&=
h_0+\sum_{p=1}^N {\Gamma_0^p\over |\xv-\av_p|}
\,,
\end{aligned}
\label{codim3harm}
\end{align}
where $\xv=(x^1,x^2,x^3)$ and $\av_p\in \mathbb{R}^3$ ($p=1,\dots,N$)
specifies the location of the codimension-3 sources where the
harmonic functions become singular. The charge vector
$\Gamma^p \equiv \{\Gamma^0_p,\Gamma^I_p,\Gamma_I^p,\Gamma_0^p\}$
carries the
charges of each source and, together with
$h\equiv
\{h^0,h^I,h_I,h_0\}$, fully determine the asymptotic
properties of the solution, namely mass, angular momenta and the moduli
at infinity.

We still have to satisfy the integrability condition
\eqref{integrability}. Because the Laplacian $\Delta$ acting on
$|\xv-\av_p|^{-1}$ gives a delta function supported at $\xv=\av_p$, the
right-hand side of \eqref{integrability} does not generally vanish.
Mathematically, this does not pose any problem for the existence of
$\omega$, although it becomes multi-valued, having a Dirac-Misner string
\cite{Misner:1963fr}.  However, the presence of a Dirac-Misner string
leads to CTCs \cite{Bena:2005va}.  Therefore, it is physically required
that the delta-function singularities be absent on the right-hand side
of \eqref{integrability}.  This condition implies the well-known
constraint \cite{Denef:2000nb}
\begin{align}
 \sum_{ q (\neq p)}
 {\bracket{ \Gamma_p , \Gamma_q } \over a_{pq}}
 =
           \bracket{h,\Gamma_p}
\qquad
\text{for each } p
\, ,
\label{balancingcond}
\end{align}
where 
$a_{pq}\equiv |\av_p-\av_q|$.

Let us see how this argument goes \cite{Bena:2005va}.  Let $B^3$ be a
small ball containing $\xv=\av_p$, and consider the integral
\begin{align}
 \int_{B^3}d^2\omega
 =\int_{B^3}d^3\xv\, 
\bracket{H,\Delta H}
,\label{intd2omega}
\end{align}
where we used \eqref{sdomega}.  The integrand on the right-hand side is
the same as the one in the integrability condition
\eqref{integrability}.  If it has a delta-function source at
$\xv=\av_p$, the integral is nonzero. On the other hand, the left-hand
side can be rewritten as
\begin{align}
 \int_{B^3}d^2\omega
 =\int_{S^2}d\omega
 =\int_{\partial S^2}\omega
\, ,\label{inted2ometa2}
\end{align}
where $S^2=\partial B^3$ and the boundary $\partial S^2$ can be taken to
be an infinitesimal circle going around the north pole, through which a
Dirac-Misner string passes. This being non-vanishing means that the
component of $\omega$ along $\partial S^2$ is finite; if we take the
Dirac-Miser string to be along the positive $z$-axis, then
$\omega_\varphi\neq 0$ where $\varphi$ is the azimuthal angle around the
$z$-axis.  Therefore, for this curve $\cC=\partial S^2$, the first term
in \eqref{omegacondCTCs} does not vanish while the second one vanishes
(note that $\cQ$ is finite as long as we are away from ${\bf x}={\bf a}_p$).  So, curve $\cC$ is a CTC\@.
Therefore, the right-hand side of the integrability condition
\eqref{integrability} must not even have delta-function singularities,
and this is what leads to the constraint \eqref{balancingcond}.

The interpretation of the singularities in the harmonic functions
\eqref{codim3harm} from a string/M-theory point of view is 
the existence of extended objects in higher dimensions.
In the string/M-theory uplift, $p$-form potentials are expressed in terms
of the harmonic functions, which allows us to establish a dictionary
between the harmonic functions and their corresponding brane
configurations \cite{Bates:2003vx}. For example, in the type IIA picture
\eqref{IIAfield}, 
the dictionary  between the singularities in the
harmonic functions and the D-brane sources is
\begin{align}
V
\leftrightarrow
\text{D6(456789)}
\, ,
\quad
\begin{array}{l}
K^1\leftrightarrow \text{D4(6789)}
\\[5pt]
K^2\leftrightarrow \text{D4(4589)}
\\[5pt]
K^3\leftrightarrow \text{D4(4567)}
\end{array}
\, ,
\quad
\begin{array}{l}
L_1\leftrightarrow \text{D2(45)}
\\[5pt]
L_2\leftrightarrow \text{D2(67)}
\\[5pt]
L_3\leftrightarrow \text{D2(89)}
\end{array}
\, ,
\quad
M \leftrightarrow \text{D0}
\, .
\label{singIIAbrn}
\end{align}
The D-branes are partially wrapped on $T^6$ and appear in 4D as
pointlike (codimension-3) objects sourcing the harmonic functions.
The components of the charge vector
$\Gamma=\{\Gamma^0,\Gamma^I,\Gamma_I,\Gamma_0\}$ are related to the
quantized D-brane numbers by
\begin{align}
 \Gamma^0= {g_s l_s \over 2}N^0,\qquad
 \Gamma^I= {g_s l_s \over 2}N^I,\qquad
 \Gamma_I= {g_s l_s \over 2}N_I,\qquad
 \Gamma_0= {g_s l_s \over 4}N_0,
 \label{charge_q'n}
\end{align}
where $N^0,N^I,N_I,N_0\in\bbZ$ (recall that the radii of the internal
torus directions have been all set to $l_s=\sqrt{\alpha'}$).  When
multiple sources are present, the harmonic solution \eqref{codim3harm}
represents a multi-center configuration of D-branes.

The harmonic solutions with codimension-3 sources have been extensively
used to describe various brane systems for various purposes.  Examples
include a 5D 3-charge black hole made  of M2(45), M2(67)
and M2(89)-branes, which is dual to the Strominger-Vafa black hole
\cite{Strominger:1996sh}; the
BMPV black hole \cite{Breckenridge:1996is};
the MSW black hole \cite{Maldacena:1997de}; the supersymmetric black
ring \cite{Elvang:2004rt, Bena:2004de, Elvang:2004ds}; multi-center
black hole/ring solutions \cite{Bates:2003vx}; and microstate geometries
\cite{Bena:2005va, Berglund:2005vb}.

One simple example is when \eqref{codim3harm} contains only one term,
namely, $N=1$.  For the generic charge vector $\Gamma\equiv \Gamma^{p=1}$,
this describes a single-center black hole in 4D which is made of D0, D2,
D4 and D6-branes.  The area-entropy of this black hole can be readily
computed to be
\begin{align}
S= \frac{\pi\sqrt{J_4(\Gamma)}}{G_4} \label{S_J4(Gamma)}
\,,
\end{align}
where the 4D Newton constant is given by $G_4=g_s^2l_s^2/8$
and $J_4(\Gamma)$ is obtained by replacing $H=\{V,K^I,L_I,M\}$ in
\eqref{J4_def0} by
$\Gamma=\{\Gamma^0,\Gamma^I,\Gamma_I,\Gamma_0\}$. Multi-center solutions
which have the same asymptotic moduli as this single-center solution and
the same total charge $\sum_p \Gamma^p=\Gamma$ can be thought of as
representing microstates/sub-ensemble of the ensemble represented by the
single-center black hole.

\bigskip
In the one-modulus class discussed in Section \ref{sec:taucases}, the
harmonic functions \eqref{codim3harm} can be rewritten in terms of the
complex harmonic function \eqref{FG_def} as
\begin{align}
 F=h_F+\sum_{p=1}^N {Q_F^p\over |{\bf x}-{\bf a}_p|},\qquad
 G=h_G+\sum_{p=1}^N {Q_G^p\over |{\bf x}-{\bf a}_p|},
\label{harm_FG}
\end{align}
where the complex quantities $(h_F,h_G)$ and $(Q_F^p,Q_G^p)$ are related
to the real quantities $h$ and $\Gamma^p$, respectively, just as $(F,G)$
are related to $H$ via \eqref{tau12harm}.  We will refer to $(Q_F,Q_G)$
as complex charges.  Using \eqref{tau12harm} and \eqref{charge_q'n}, 
we can see that 
they are related to quantized charges by
\begin{align}
\begin{split}
  Q_F&={g_s l_s\over 2}(N^3+iN_1),\qquad
 Q_G={g_s l_s\over 2}(N^0-iN^1),
 \\[1ex]
 N^1&=N^2,\qquad N_1=N_2,\qquad N^0=N_3,\qquad N^3=-N_0.
\end{split}
\label{QFQG_q'n}
\end{align}
The black-hole entropy \eqref{S_J4(Gamma)} can be written as
\begin{align}
 S={8\pi\,|\!\Im(Q_F \Qb_G)|\over g_s^2 l_s^2}
 =2\pi\,|N^3 N^1+N_1 N^0|\,.
\label{S_QFQG}
\end{align}

\subsection{Codimension-2 solutions}

\bigskip
\subsubsection*{Codimension-2 sources are inevitable}

In addition to codimension-3 sources, the harmonic solutions can also describe
codimension-2 sources.  Actually, codimension-2 sources are \emph{not an
option but a must}; codimension-3 sources are insufficient because they
can spontaneously polarize into codimension-2 sources by the
\emph{supertube transition} \cite{Mateos:2001qs}.  The supertube
transition is a spontaneous polarization phenomenon that a certain pair
of species of branes --- specifically, any 1/4-BPS 2-charge system ---
undergo.  In this transition, the original branes polarize into a new
dipole charge, which has one less codimension and extends along a closed
curve transverse to the worldvolume of the original branes.  This new
configuration represents a genuine BPS bound state of the 2-charge
system \cite[Sec.~3.1]{Mathur:2005zp}.
The supertube transition may seem similar to the Myers effect
\cite{Myers:1999ps}, but it is different; the Myers effect takes place
only in the presence of an external field, whereas the supertube
transition occurs spontaneously, by the dynamics of the system itself.

The system described by codimension-3 harmonic solutions involves various
D-branes as we saw in \eqref{singIIAbrn}.  These D-branes can undergo
supertube transitions into codimension-2 branes, which act as
codimension-2 sources in the harmonic function.  Therefore,
codimension-2 solutions are in the same moduli space of physical
configurations as codimension-3 solutions, and consequently \emph{must}
be considered if one wants to understand the physics of the D-brane
system.

In particular, supertubes are known to be important for BPS microstate
counting of black holes because of the entropy enhancement phenomenon
\cite{Bena:2008nh,Bena:2008dw,deBoer:2009un, Bena:2010gg}.
So, the supertubes realized as
codimension-2 sources in the harmonic functions must play a crucial
role in the black hole microstate geometry program, as first argued in
\cite{deBoer:2010ud, deBoer:2012ma}.  The codimension-2 brane produced
by the supertube transition can generically be non-geometric, having
non-geometric U-duality twists around it.

A prototypical example of the supertube transition \cite{Mateos:2001qs}
can be represented as
\begin{align}
\text{D0}+ \text{F1}(1)
\to
\text{d2}(\lambda 1)
\, .
\label{D0+F1->d2}
\end{align}
This diagram means that the 2-charge system of D0-branes and F1-strings
has undergone a supertube transition and polarized into a D2-brane
along an arbitrary closed curve parametrized by $\lambda$.  The object
on the right-hand side is written in lowercase to denote that it is a
dipole charge. In this case, as the D2 is along a closed curve, there is
no net charge but a D2 dipole charge.  The original D0 and F1 charges
are dissolved into the D2 worldvolume as magnetic and electric fluxes.
The Poynting momentum due to the fluxes generates the centrifugal force
that prevents the arbitrary shape from collapsing.

Upon duality transformations of the process \eqref{D0+F1->d2}, other
possible supertube transitions can be found.  For example,
\begin{align}
\begin{array}{l@{~}c@{~}l@{~}c@{~}l}
 \text{D0}      &+& \text{D4}(4567)  &\quad\to\quad&  \text{ns5}(\lambda 4567)\, ,\\[.5ex]
 \text{D4}(4589)&+& \text{D4}(6789)  &\quad\to\quad& 5^2_2(\lambda 4567;89)\, ,\\[.5ex]
 \text{D2}(45)  &+& \text{D2}(67)    &\quad\to\quad&  \text{ns5}(\lambda 4567)\, ,\\[.5ex]
 \text{D2}(89)  &+& \text{D6}(456789)&\quad\to\quad& 5^2_2(\lambda 4567;89)\, .\\
\end{array}
\label{codim2supertubes}
\end{align}
This means that the ordinary branes on the left-hand side can polarize
into codimension-2 branes, including the exotic branes such as the
$5^2_2$-brane.\footnote{For a review on exotic branes and a further
analysis of supertube transitions involving them, see
\cite{deBoer:2012ma}. We discuss a $\text{D2}+\text{D6}\to 5^2_2$
transition  in Appendix \ref{app:D2+D6->522}.}
Note in particular that the D-branes appearing on the left-hand side are
the ones that appear in the brane-harmonic function dictionary
\eqref{singIIAbrn}.  So, the dictionary is insufficient and must
be extended to include codimension-2 branes that the codimension-3
D-branes can polarize into.  Because we solved the BPS equations and
obtained harmonic solutions without specifying the co-dimensionality of
the sources, the codimension-2 supertubes on the right-hand side of
\eqref{codim2supertubes} must be describable in terms of the same
harmonic solutions, just by allowing for codimension-2 singularities.
The formulas for the M-theory/IIA uplift also remain valid.

\subsubsection*{Examples of  codimension-2 solutions}

Let us study some codimension-2 solutions that are given in terms of the harmonic solutions. From \eqref{codim2supertubes}
let us consider the following process:
\begin{align}
\text{D2}(45) + \text{D2}(67) 
\to 
\text{ns5}(\lambda 4567)
\, .
\label{D2+D2->ns5}
\end{align}
It was shown in \cite{Park:2015gka} that the codimension-2 ns5 supertube
on the right-hand side can be described within harmonic solutions by the
following harmonic functions
\begin{align}
\label{D2+D2->ns5_harm} 
\begin{aligned}
V
&=
1
\, ,
\quad
&
K^1
&=
0 
\, , 
\quad
&
K^2
&=
0
\, , 
\quad
&
K^3
&=
\gamma
\, , &&
\\
&&
L_1
&=
f_2 
&
\quad
L_2
&=
f_1
\, , 
&
\quad
L_3
&=
1
\, , 
\quad
&
M
&=
-\frac \gamma2
\, ,
\end{aligned}
\end{align}\
where
\begin{align}
f_1
&=
1
+\frac{Q_1}{L}\int_0^L \frac{d\lambda}{|\xv- \Fv(\lambda)|}
\, ,
&
f_2
&=
1
+\frac{Q_1}{L}\int_0^L \frac{|\dot \Fv(\lambda)|^2 \, d\lambda}{|\xv- \Fv(\lambda)|}
\, .
\label{f1,f2_def}
\end{align}
The supertube lies along the closed curve $\xv=\Fv(\lambda)$, where
$F_i(\lambda)$ ($i=1,2,3$) are arbitrary functions satisfying
$F_i(\lambda+L)=F_i(\lambda)$.  $Q_1$ is the D2(67)-brane charge, while
the D2(45)-brane charge is given by $Q_2={Q_1\over L}\int_0^L
|\dot{\Fv}(\lambda)|^2\,d\lambda.$ The integrals in \eqref{f1,f2_def}
arise as a consequence of these charges being dissolved along the
worldvolume of the supertube. For expressions of $L,Q_1,Q_2$ in terms of
microscopic quantities, see \cite{Park:2015gka}.  $\gamma$ is a harmonic
scalar function defined through the equation
\begin{align}
d\alpha
=
*_3 d\gamma
\, ,
\qquad\qquad
\alpha
=
\frac{Q_1}{L}\int_0^L \frac{\dot F_i(\lambda)\, d\lambda}{|\xv-\Fv(\lambda)|}\,dx^i
\, .\label{alpha_gamma}
\end{align}
Even though the 1-form $\alpha$ is single-valued, $\gamma$ is
multi-valued and has monodromy as we go once around the supertube
\cite{Park:2015gka}:
\begin{align}
 \gamma \to \gamma + 1.
\label{gamma_monodromy}
\end{align}
The integrability condition \eqref{integrability} is satisfied without
any delta-function singularity along the profile, because $\Delta
\gamma=0$ without any singular contribution on the profile
\cite{Park:2015gka}.
Other data of the harmonic solutions are
\begin{align}
Z_I 
&=
( f_2, f_1, 1)
\, ,
&
\mu 
&=
0
\, ,
&
\omega
&=
-\alpha
\, ,
&
\xi^I
&=
( 0, 0,-\alpha )
\, .
\end{align}

The charge content of the solution can be easily read off from the
harmonic functions.  The original codimension-3 charges for D2(45) and
D2(67) are encoded in $L_1$ and $L_2$ by the dictionary
\eqref{singIIAbrn}.    From
\eqref{f1,f2_def}, we see that these charges are distributed along the
profile $\xv=\Fv(\lambda)$ with densities $Q_1/L$ and
$(Q_1/L)|\dot\Fv|^2$, respectively.  On the other hand, the NS5 charge
is encoded in the monodromy.  Eq.~\eqref{gamma_monodromy} means the
following monodromy around the supertube:
\begin{align}
 \begin{pmatrix}  K^3\\ V \end{pmatrix}
 =
 \begin{pmatrix}  \gamma \\ 1 \end{pmatrix}
 \to
 \begin{pmatrix}  \gamma +1\\ 1 \end{pmatrix}
 =
 \begin{pmatrix}  1 & 1\\ 0 & 1 \end{pmatrix}
 \begin{pmatrix}  K^3\\ V \end{pmatrix}.
\end{align}
From \eqref{SL2transf_harmfunc}, \eqref{SL2doublets}, this means that we
have the following $\text{SL}(2,\bbZ)_3$ monodromy:
\begin{align}
 M_3=\begin{pmatrix}  1 & 1\\ 0 & 1 \end{pmatrix}\in \text{SL}(2,\bbZ)_3.
\end{align}
One can also see this from the K\"ahler moduli,
\begin{align}
\tau^1&=i\sqrt{\frac{f_1}{f_2}}\, ,\qquad
\tau^2=i\sqrt{\frac{f_2}{f_1}}\, ,\qquad
\tau^3=\gamma +i\sqrt{{f_1}{f_2}}\, .\label{tauI_D2D2ns5}
\end{align}
We see that, as we go once around the supertube, $\tau^{1,2}$ are
single-valued whereas $\tau^3$ has the monodromy
\begin{align}
\tau^3 \to \tau^3+1
\, .
\end{align}
Because $\Re\tau^3=B_{89}$, this monodromy implies that there is an
NS5-brane along the closed curve.

%


One can consider other configurations involving codimension-2 branes.
In Appendix \ref{app:D2+D6->522}, we discuss the D2(89)+D6(456789)$\to
5^2_2(\lambda 4567;89)$ supertube, which is the last entry in
\eqref{codim2supertubes} and was studied in \cite{Park:2015gka}.  

In the special case where $|\dot{\bf F}|=1$, we have $f_1=f_2\equiv f$ and
therefore $\tau^1=\tau^2=i$ as we can see from \eqref{tauI_D2D2ns5}.
This case belongs to the one-modulus class discussed in Section
\ref{sec:taucases}, with the complex harmonic functions
\begin{align}
 F=\gamma+if,\qquad G=1.\label{FG_D2D2ns5}
\end{align}
This setup is simple but still non-trivial enough to
include interesting physical situations such as the $\text{D2}(45) +
\text{D2}(67) \to \text{ns5}(\lambda 4567)$ supertube. It can also
describe the D2(89)+D6(456789)$\to 5^2_2(\lambda 4567;89)$ supertube
discussed in Appendix \ref{app:D2+D6->522}\@. We will use this setup to
construct a non-Abelian supertube configuration involving both
$\text{D2} + \text{D2} \to \text{ns5}$ and $\text{D2}+\text{D6}\to
5^2_2$ supertubes.

In the above we discussed configurations just with codimension-3 sources
or just with codimension-2 sources.  One can also consider a mixed
configuration in which a codimension-3 source and a codimension-2 source
coexist \cite{Park:2015gka}.

\subsubsection*{General remarks on codimension-2 solutions}

For the codimension-3 case, we could show the direct connection between
the presence of delta-function sources on the right-hand side of
equation \eqref{integrability} and the existence of CTCs\@.  We can
follow the same line of logic for the codimension-2 case, but 
the conclusion is that there is no such direct connection.

In \eqref{intd2omega}, we had an integral over a small ball $B_3$
containing a point where there is a possible delta function.  In the
codimension-2 case, delta-function singularities are expected to be
along a curve on which a source lies, and there is a Dirac-Misner
``sheet'' ending on that curve.  Let us consider an integral over a very
thin filled tube $T^3$ containing a piece of such a curve. Now we
rewrite the integral as we did in \eqref{inted2ometa2}.  Instead of
$S^2=\p B^3$, we have a cylinder $C^2=\p T^3$, where we can ignore the
top and bottom bases for a very thin tube.  As the boundary of the
cylinder, $\p C^2$, we take two lines that go along the curve in
opposite directions.  The Dirac-Misner sheet goes between the two lines.
Then the integral is basically equal to the jump across the Dirac-Miner
sheet in the component of $\omega$ along the curve. Let us denote it by
${\mathit \Delta}\omega_\parallel$. Then, the integral is equal to
$l{\mathit \Delta}\omega_\parallel$, where $l$ is the length of the
tube.  On the other hand, the same integral is equal to $l\sigma$, where
$\sigma$ is the local density of the delta-function source along the
curve.  Equating the two, we obtain
\begin{align}
 {\mathit \Delta}\omega_\parallel=\sigma.
\label{Domega=sigma}
\end{align}
Namely, the jump in $\omega$ along the curve is given by the density of
delta-function sources.

However, this does not give the behavior of $\omega$ itself, which is
necessary for evaluating \eqref{omegacondCTCs} and study the presence of
CTCs.  So, the argument that worked for codimension 3 does not apply to
codimension 2.  It must be some other singular behavior of the
harmonic functions, not just the delta-function source, that one must
study to investigate the no-CTC condition.  We do not pursue that in
this paper.  Instead, we will study \eqref{omegacondCTCs} for specific
explicit metrics in the presence of codimension-2 sources.

\bigskip
For codimension-3 sources, construction of general multi-center
solutions is straightforward because of ``linearity'': one can simply
add the poles representing different codimension-3 sources, as we did in
\eqref{codim3harm}.  However, in contrast, construction of general
solutions with multiple codimension-2 sources is less straightforward.
This is because linearity is lost if there are multiple codimension-2
objects whose monodromy matrices do not commute, in other words, if the
monodromies are non-Abelian.
Indeed, the explicit construction of solutions with multiple
codimension-2 supertubes thus far \cite{Park:2015gka} is restricted to
the case where (i) all supertubes have the same monodromy, or (ii)
different supertubes have different monodromies but they all commute
with each other.  In either case, the monodromies are Abelian.  In such
cases, linearity still holds and the corresponding harmonic functions
can be obtained by adding harmonic functions for each
supertube.\footnote{See Footnote \ref{ftnt:linearity}.}
In the next section, we will construct a configuration of two supertubes
with non-Abelian monodromies in a certain limit.

Although we have only discussed sources with codimension 3 and 2, it is also
possible to consider sources with codimension 1.  Such a source
represents a domain wall that connects spaces with different values of
spacetime-filling fluxes, just like a D8-brane in 10D connects
spacetimes with different values of the RR flux 10-form. Including
codimension-1 sources should lead to a wide range of physical
configurations which have been little studied. It would be very
interesting to include them in the harmonic solutions and explore the
physical implications of solutions with codimension 3, 2, and 1 sources.


\section{Explicit construction of non-Abelian supertubes}
\label{sec:examnonAbel}



\subsection{Non-Abelian supertubes}
\label{sec:nonAbel}



In the previous section, we saw that harmonic solutions can describe BPS
configurations of codimension-2 supertubes.  A codimension-2 supertube
has a non-trivial U-duality monodromy around it, which can be
represented by a monodromy matrix $M$.
If multiple co\-di\-men\-sion-2 supertubes are present and the $i$-th
supertube has a monodromy matrix $M_i$ then, in general,
the monodromies of different supertubes do not commute,
$[M_i,M_j]\neq 0$ for some pair $(i,j)$, namely, the monodromies are \emph{non-Abelian}. In
this section, we show, for the first time, that such a
non-Abelian configuration of supertubes is indeed possible.

\begin{figure}[h]
\begin{quote}
\capstart
\begin{center}
 \includegraphics[width=0.45\textwidth]{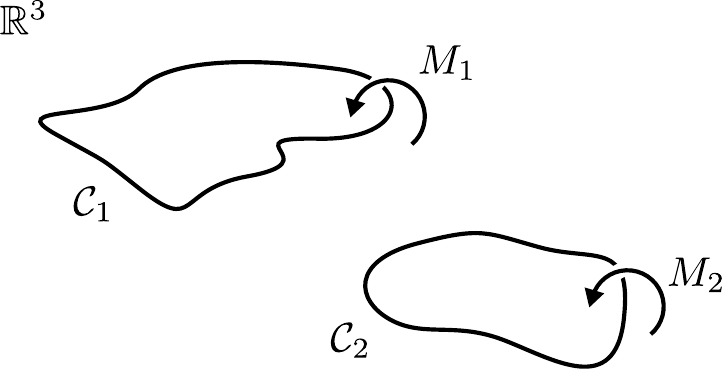}
 \caption{\sl A non-Abelian configuration of two supertubes. The monodromy
 matrices $M_1,M_2$ of the two supertubes do not commute,
 $[M_1,M_2]\ne0$.}  \label{fig:gennonAbel}
\end{center}
\end{quote}
\end{figure}

We will focus on configurations in which only one modulus $\tau^3\equiv
\tau$ is non-trivial and has $\text{SL}(2,\bbZ)$ monodromies.  As discussed in Section \ref{sec:taucases}, in this situation, only four harmonic
functions are independent \eqref{tau12iharmcond}, which can be combined into two complex
harmonic functions $F,G$.  In terms of them, the modulus $\tau$ can be
written as
\begin{equation}
\tau=\frac{F}{G}\,.\label{tau=F/G}
\end{equation}
The simplest non-Abelian configuration is one with two supertubes.  As
we go around the $i$-th supertube, the harmonic functions transform as
\begin{equation}\label{FGmon}
\begin{pmatrix}
F \\ G
\end{pmatrix}
\to
M_i \begin{pmatrix}
F \\ G
\end{pmatrix}\,,\qquad
M_i\in \text{SL}(2,\bbZ)\,,\qquad
i=1,2.
\end{equation}
We require that the monodromies be non-Abelian,
\begin{align}
 [M_1,M_2]\neq 0.\label{[M1,M2]neq0}
\end{align}
See Figure~\ref{fig:gennonAbel} for a pictorial description of 
such a 2-supertube configuration.

Specifically, we will consider a two-supertube configuration with the
following monodromies:
\begin{align}
\label{M1M2}
M_1&=\begin{pmatrix}\phantom{-}1 & 0 \\ -2 & 1\end{pmatrix}\,,\qquad 
M_2=\begin{pmatrix}\phantom{-}3 & \phantom{-}2 \\ -2 & -1\end{pmatrix}\,.
\end{align}
These clearly give a non-Abelian pair of monodromies
satisfying~\eqref{[M1,M2]neq0}.
As we will discuss later in this section, this choice is motivated by
the solution to a similar monodromy problem discussed in the $\LG{SU}(2)$
Seiberg-Witten theory \cite{Seiberg:1994rs}. If we go
around the two supertubes, the total monodromy is
\begin{align}
M=M_2 M_1=\begin{pmatrix}-1 & \phantom{-}2 \\ \phantom{-}0 & -1\end{pmatrix}.
\label{Mtotal}
\end{align}

If one is far away from the supertubes, none of the monodromies
of the supertubes are visible and the configuration looks like that of a
single-center codimension-3 solution.  From the $|\xv|\to\infty$ behavior of the
harmonic functions, we can read off the charges of the single-center
solution. We will find that the charges are those of a 4-charge black
hole in four dimensions with a finite horizon. In other words, seen from
a large distance, our configuration looks like an ordinary 4-charge
black hole without any monodromic structure.  However, as one approaches
it, the topology of the supertubes becomes distinguishable and discovers that the spacetime
has non-trivial non-Abelian monodromies.




\subsection{Strategy}
\label{sec:strategy}

The problem that we should attack in principle is the following.  We
first specify two closed curves $\cC_1,\cC_2$ in $\bbR^3$ along which
the two supertubes lie, such as the ones in Figure~\ref{fig:gennonAbel}.
Then we must find a pair of harmonic functions $(F,G)$ which, as
we go around curve $\cC_i$, undergoes the monodromy transformation
\eqref{FGmon} with the monodromy matrix $M_i$ given in~\eqref{M1M2}.
If we can find such pair $(F,G)$, then the
configuration exists.

Although this is a mathematically well-posed problem, explicitly
carrying it out for general shapes of supertubes is technically
challenging.
Instead, our strategy here is to take a particularly simple
configuration for the two supertubes and further take a limit in which
the problem of finding the solution becomes simple and tractable but  is
still non-trivial.  This is sufficient for the purpose of proving the
existence of a configuration of non-Abelian supertubes.

\begin{figure}[h]
\begin{quote}
 \capstart
 \begin{center}
 \includegraphics[width=0.90\textwidth]{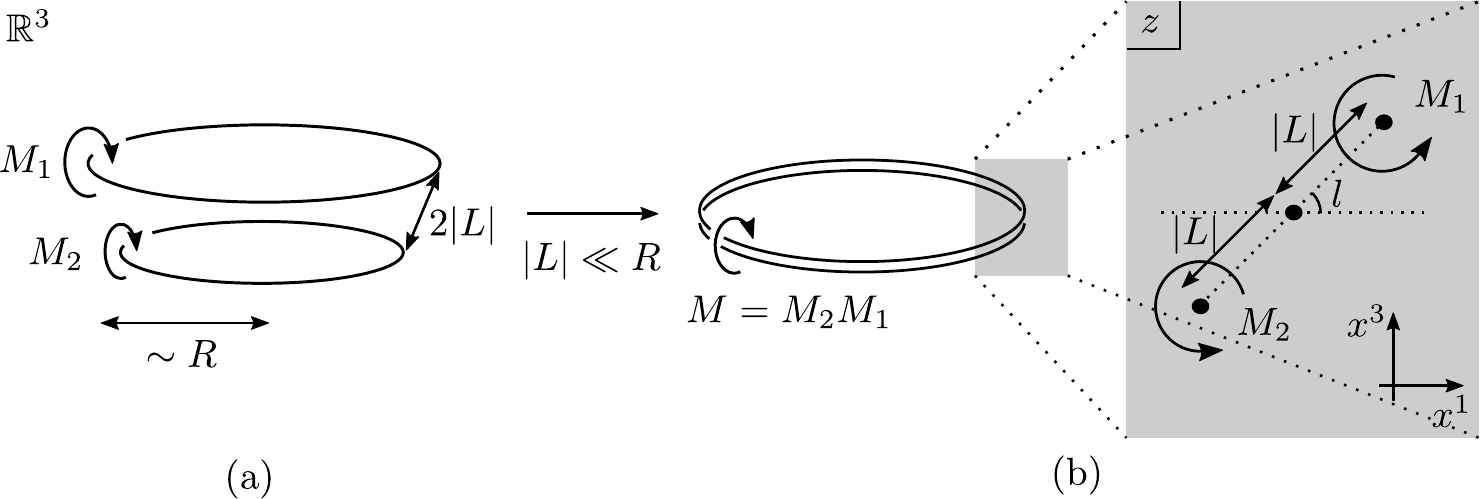}
 \caption{{\sl (a)~A configuration of two circular supertubes sharing the
 axis. (b)~The configuration in the colliding limit, $\abs{L}\ll R$. In this
 limit, we can study the problem in two different regimes, the near and
 far regions. In the near region, the system becomes 2-dimensional but we
 must consider two separate monodromies $M_1,M_2$ of two supertubes.  In
 the far region, the system remains 3-dimensional but there is only one
 tube with monodromy $M=M_2M_1$.}}  \label{fig:colliding}
 \end{center}
\end{quote}
\end{figure}

Specifically, we assume that the two tubes are circular and share the
axis (so that the configuration is axisymmetric). The two tubes have
almost identical radius $R>0$ and are very close to each other,
separated by distance $2|L|$; see Figure~\ref{fig:colliding}(a). More
precisely, in equations, the location of supertubes 1 and 2 is specified
as follows:
\begin{align}
\label{tubes_pos}
\begin{split}
  \text{Supertube 1:}\qquad  & (x^1)^2+(x^2)^2=(R+|L|\cos l)^2,\quad x^3=+|L|\sin l,\\
 \text{Supertube 2:}\qquad  & (x^1)^2+(x^2)^2=(R-|L|\cos l)^2,\quad x^3=-|L|\sin l,
\end{split}
\end{align}
where $l$ is the angle between the two tubes relative to the $x^1$-$x^2$
plane; for example, $l=0$ if they are concentric.  We study this system
in the \emph{colliding limit},
\begin{align}
 |L|\ll R. \label{colliding_lim}
\end{align}
In this
limit, we can break down the problem into two regimes, depending on the
distance $d$ from an observer to the supertubes, as follows:
\begin{itemize}
 \item[(i)] {\it The near region, $d\sim |L|\ll R$.} 

	    \nopagebreak
	    In this region, the two supertubes can be regarded as
	    infinite straight lines and we can forget the direction
	    along them.  Therefore, the system can effectively be
	    treated as 2-dimensional.  By symmetry, we can
	    zoom in onto the region near the point
	    $(x^1,x^2,x^3)=(R,0,0)$ without loss of generality, and
	    identify the $z$-plane with a small piece of the $x^1$-$x^3$
	    plane near that point with the relation
	    \begin{align}
	     z=(x^1-R)+ix^3,\qquad\qquad
	     |x^1-R|,|x^3|\sim |L| \ll R.\label{z_and_R3}
	    \end{align}
	    On the $z$-plane, the two supertubes are located at $z=L$
	    and $z=-L$, where we defined
	    \begin{align}
	     L=|L|e^{il}.\label{L=|L|e^il}
	    \end{align}

	    So, the problem reduces to that of finding on the $z$-plane
	    a pair of 2D harmonic functions $(F,G)$ that
	    has non-trivial monodromies $M_1,M_2$ given in \eqref{M1M2}
	    around $z=\pm L$.  See Figure \ref{fig:colliding}(b).
 \item[(ii)] {\it The far region, $|L|\ll R\sim d$}.

	    \nopagebreak
	     In this region, the two supertubes cannot be resolved and
	    we effectively have only one supertube sitting at
	     \begin{align}
 	    (x^1)^2+(x^2)^2=R^2, \qquad x^3=0
 	    , ,
	      \label{tube_pos_far}
	     \end{align}	     
	     with the combined monodromy
	    $M=M_2M_1$ given in \eqref{Mtotal}.  So, the
	    problem reduces to that of finding 3D harmonic functions
	    $(F,G)$ with the monodromy $M$ around one circular
	    supertube.
\end{itemize}

After finding the expressions for the harmonic functions $(F,G)$ in 
regions (i) and (ii), we must connect them in the intermediate region, $|L|\ll
d\ll R$, in order to show the existence of $(F,G)$ defined in the entire
space.  Namely, we must match the large-$|z|$ behavior of the
near-region solution smoothly onto the near-ring (i.e.,
$(x^1,x^2,x^3)\to (R,0,0)$) behavior of the far-region solution.

This matching can be done order by order and the harmonic function in
the entire space can be reconstructed to any order in perturbative
expansion.  To see exactly how this works in practice, let us study a
toy example in which we can work out the matching procedure in detail.

\subsubsection*{A toy model for the matching procedure}

As a simpler physical problem in which there are two very different
scales $|L|$ and $R$ with $|L|\ll R$, let us consider the following problem.
In three dimensions, we would like to find the field configuration
sourced by two point-like charges at $\xv=\pm\Lv\equiv (0,0,\pm |L|)$ with
charge $Q_\pm$.  Assume that the field $H$ is governed by the Helmholtz
equation
\begin{equation}
\left(
\Delta
	- \frac{1}{R^2}
	\right) H 
=
0
\, .
\label{Helmholtz}
\end{equation}
Of course, for this problem, we know the exact answer:
\begin{equation}
\label{KGsol0}
H
=
\frac{Q_+ e^{-\frac{\abs{\xv- \Lv}}{R}}}{|\xv - \Lv|}
+ \frac{Q_- e^{-\frac{\abs{\xv+ \Lv}}{R}}}{|\xv + \Lv|}
\, .
\end{equation}
However, let us try here to recover this expression by working in the
``near region'' $|\xv|\sim |L|\ll R$ and in the ``far region'' $|L|\ll R\sim
|\xv|$ separately, and finally matching the expressions in the
intermediate region connecting the two.

In the near region $|\xv|\sim |L|\ll R$, we can ignore the $R$ dependence
in \eqref{Helmholtz}.  Therefore, the expression in the near region is
\begin{equation}
H
=\frac{Q_+}{|\xv - \Lv|}+\frac{Q_-}{|\xv + \Lv|}
\, .
\end{equation}
Let $(r,\theta,\varphi)$ be the spherical polar coordinates for $\bbR^3$.  If
we increase $r$, still staying inside the near region, we can do a small
$|L|\over r$ expansion of this and obtain
\begin{equation}\label{kahc20Apr17}
H
=
\frac{Q_+ + Q_-}{r} +\frac{(Q_+ - Q_-)|L|\cos\theta}{r^2}+ 
\mathcal{O}\left({|L|^2\over r^3}\right)
\, ,
\end{equation}
which corresponds to the standard multipole expansion. We would like to
find how this multipole expansion matches onto the one in the
far region.

To be able to do the matching, there must be an intermediate region where the expansion~\eqref{kahc20Apr17} is correct.  To understand what
this means, let us make the scaling for the intermediate region, $|L|\ll r \ll R$, more precise by
setting
\begin{equation}
\frac{r}{R}\sim \epsilon
\, ,
\qquad
\frac{|L|}{r}\sim \delta
\, ,
\end{equation}
where $\epsilon,\delta\ll1$. If we are to keep $r$ finite,
the replacement
\begin{equation}
R \to R\epsilon^{-1}
\, ,
\qquad
|L| \to |L|\delta
\, ,
\end{equation}
will keep track of the order of expansion.  If we do this replacement in
the exact expression~\eqref{KGsol0} and expand it in powers of
$\epsilon$ and $\delta$, we obtain
\begin{align}
H
 &=
 \left[\frac{Q_+ + Q_-}{r}
 +\frac{(Q_+ - Q_-)|L|\cos\theta}{r^2} \delta 
 +\mathcal{O}(\delta^2)\right]
 -\frac{(Q_+ + Q_-)\epsilon}{R}
\notag\\
 &\qquad
 +   \left[\frac{(Q_+ + Q_-)r}{2 R^2}
 -\frac{(Q_+ - Q_-)|L| \cos\theta }{2R^2}\delta
 +\mathcal{O}(\delta^2)\right]\epsilon ^2
 +\mathcal{O}(\epsilon^3)
 \, .
\label{jrrd14Jul17}
\end{align}
If we make $\epsilon$ small enough so that
only the $\mathcal{O}(\epsilon^0)$ terms remain, then this reproduces the near-region
expansion \eqref{kahc20Apr17}.  Therefore, the correct procedure is:
take $\epsilon\to 0$ first, and then match the $\delta$ expansion.  In
other words, take $R\to \infty$ first, and then match the small $|L|\over r$
expansion.

With this mind, let us go to the far region.  Here, the two charges
cannot be resolved and the function $H$ can be singular only at
$r=0$.  The instruction is: find solutions of the Helmholtz equation
such that their $R\to 0$ limit reproduces \eqref{kahc20Apr17}, term by
term in the $|L|\over r$ expansion.  First, 
\begin{align}
 (Q_+ + Q_-){e^{-{r\over R}}\over r}
\end{align}
is clearly an exact solution with a singularity at $r=0$.  If we take
$R\to\infty$, this gives $r^{-1}$, which reproduces the first term in
\eqref{kahc20Apr17}.  Next,
\begin{equation}\label{KGsol2}
 (Q_+ - Q_-)\,|L|\,e^{-{r\over R}}
  \left({1\over r^2}+{1\over Rr}\right)\cos\theta
\end{equation}
is an exact solution and its $R\to\infty$ limit reproduces the second
term in \eqref{kahc20Apr17}.  So, up to this order,
the far-region solution which reproduces \eqref{kahc20Apr17} is
\begin{equation}
\label{mepq19Aug17}
 H
 =
 {(Q_+ + Q_-)\,e^{-{r\over R}}\over r}
 + (Q_+ - Q_-)\,|L|\,e^{-{r\over R}}
 \left({1\over r^2}+{1\over Rr}\right)\cos\theta
 \, + \cO\left({|L|^2\over r^3}\right).
\end{equation}
It is clear that we can keep going with this procedure to find the
far-region solution that reproduces the expansion \eqref{kahc20Apr17} to
an arbitrarily high order, upon taking the $R\to\infty$ limit.  In
principle, if we can sum this expansion to all orders, we can recover
the exact expression~\eqref{KGsol0} with singular sources at ${\bf
x}=\pm{\bf L}$.  However, at any finite order, the perturbative
expression~\eqref{mepq19Aug17} has a singularity only at $r=0$; namely,
some features of the exact solution can be seen only after carrying out
the infinite sum, which is a limitation of the method of matching
expansion.

Below, we will use the exactly same matching procedure to find the
harmonic functions describing a configuration of non-Abelian supertubes.


\subsection{The near region}
\label{sec:near}

Now with the colliding limit and the matching procedure
understood, let us construct the solution starting from the
near-region side.

\subsubsection*{Some general statements}

As we mentioned before, in the near region, we can regard the round
supertubes as parallel, infinite straight lines. Forgetting about the
direction along the tubes, the problem reduces to the one on the
$z$-plane defined in \eqref{z_and_R3}.  A harmonic function in 2D can be
written as the sum of holomorphic and anti-holomorphic functions.  In
the present case, this means that $F,G$ are both written as a sum of
holomorphic and anti-holomorphic functions.

Let us further assume that $F$ and $G$ are purely holomorphic:
\begin{align}
 F=F(z),\qquad G=G(z).
\end{align}
This is equivalent to assuming that $\tau=F/G$ is holomorphic.  In this
case, we can solve \eqref{domega_FG} to find $\omega$ explicitly.  If we set
\begin{align}
 \omega=\omega_2 dx^2+\omega_z dz+\omega_\zb d\zb,
\end{align}
where $\omega_z$, $\omega_\zb$ and $\omega_2$ are independent of $x^2$,
then
\begin{align}
 \omega_2=-\Im(F\Gb)+C,\qquad \p\omega_\zb-\pb\omega_z=0
\label{omega_2D}
\end{align}
where $C$ is a constant.

The above $\omega_2$ is $\text{SL}(2,\bbZ)$ invariant because
\begin{align}
\begin{pmatrix} \alpha&\beta\\ \gamma&\delta\end{pmatrix}:\quad
 \Im(F\Gb)\to \Im[(\alpha F+\beta G)(\gamma\Fb+\delta\Gb)]
  =\Im[ (\alpha\delta-\beta\gamma) F\Gb ]  =\Im(F\Gb ),
\end{align}
for $\alpha\delta-\beta\gamma=1$. Therefore, even if there is a singularity around which
there is an $\text{SL}(2,\bbZ)$ monodromy and $(F,G)$ are multi-valued,
$\omega_2$ is always single-valued.  By \eqref{Domega=sigma}, this means
that the integrability condition \eqref{integrability} is satisfied
without delta-function singularities along the supertube.

The constant $C$ and functions
$\omega_z,\omega_\zb$ must ultimately be fixed by extending the
near-region solution to the far-region solution and requiring that
$\omega$ be regular everywhere and vanish at 3D infinity.
In the present case, we will find that $\omega$ in the far region has a
non-vanishing component only in the direction along the supertube.
Therefore, we set $\omega_z=\omega_\zb=0$.  On the other hand, the
constant $C$ cannot be fixed unless we have an exact solution (we only
have a perturbative solution in the present paper).

When there is a supertube, the direction along its profile is a dangerous
direction where there can be CTCs \cite{Emparan:2001ux, Mateos:2001pi}.
This is the $x^2$ direction in the present case and the $22$ component of
the metric which is, e.g., from \eqref{IIAfield},
\begin{align}
 g_{22}\propto -\omega_2^2+\cQ
 =-[-\Im(F\bar{G})+C]^2+[\Im(F\bar{G})]^2
 =C[2\Im(F\bar{G})-C].
\label{g22}
\end{align}
From \eqref{tau12CTCs}, $\Im(F\bar{G})\ge 0$.  So, for \eqref{g22} not
to be negative, the constant $C$ must be in the following range:
\begin{align}
 0\le C \le 2\min[\Im(F\bar{G})].\label{C_cond}
\end{align}
This does not have to hold up to $z=\infty$.  It only has to hold up to
some value of $|z|$ above which the 2D approximation breaks down.

\subsubsection*{The solution}

On the $z$-plane, we would like to construct a pair of harmonic
functions $(F,G)$ that has non-trivial non-Abelian monodromy
\eqref{FGmon} around some singular points.  In doing that, we must
require that the imaginary part of $\tau={F/ G}$ be always positive,
because of the condition \eqref{tau12CTCs}.  There are
many such possibilities, but in this paper we will take the pair of
holomorphic functions that appeared in the solution of $d=4,\cN=2$
supersymmetric gauge theory by Seiberg and Witten \cite{Seiberg:1994rs},
because it is a fundamental example of configurations with non-Abelian
monodromies.

\begin{figure}[h]
\begin{quote}
 \capstart
 \begin{center}
 \includegraphics[height=0.24\textwidth]{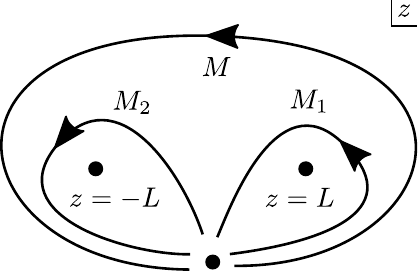}
 \caption{\sl
 The monodromy structure in the near region.  At $z=\pm L$ we have
 singularities corresponding to the position of the supertubes. When
 going around one of them, $(F,G)$ gets transformed by $M_i$. Going
 around both of them induces a monodromy transformation $M=M_2M_1$.}
 \label{fig:2dplane}
 \end{center}
\end{quote}
\end{figure}

The original work of Seiberg and Witten was about the exact
determination of the low-energy effective theory of $\mathcal{N}=2$ pure
$\LG{SU}(2)$ gauge theory.  At low energy, the theory has a Coulomb
moduli space parametrized by the vacuum expectation value of the vector
multiplet scalar, $z=\ev{\tr\phi^2}\in\bbC$.  At point $z$ on the moduli
space, one has a pair of holomorphic functions $(a_D(z),a(z))$ which represent
the mass of the magnetic monopole and the electron at that point.  In
terms of them, the low-energy coupling constant, $\tau(z)$, is expressed
as
\begin{equation}\label{def:tau}
\tau(z)=\frac{da_D}{da}=\frac{a_D'(z)}{a'(z)}
\,.
\end{equation}
The theory has an $\text{SL}(2,\bbZ)$ duality group which changes the
coupling constant $\tau$ and acts non-trivially on the spectrum of
dyons. More specifically, under $\text{SL}(2,\bbZ)$, the pair $(a_D,a)$
transforms as a doublet and $\tau$ undergoes linear fractional
transformation.  The moduli space has three singularities at $z=\pm
L,\infty$ around which there are non-trivial monodromies of the
$\text{SL}(2,\bbZ)$ duality.  The one at $z=L$ is due to the magnetic
monopole becoming massless and the monodromy around it is given by $M_1$
in \eqref{M1M2}.  On the other hand, the one at $z=-L$ is due to the
$(1,1)$ dyon getting massless and the monodromy is given by $M_2$ in
\eqref{M1M2}.  Finally, the one at $z=\infty$ is due to asymptotic
freedom and the monodromy is given by $M$ in \eqref{Mtotal}.  See
Figure~\ref{fig:2dplane} for the monodromy structure of the moduli
space.

One sees that this theory has everything we need.  We identify the
$\text{SL}(2,\bbZ)$ duality group on the gauge theory side with the
$\text{SL}(2,\bbZ)_3$ U-duality group on the supertube side, the modulus
$z$ with the $z$ coordinate of the near region, the mass parameters
$(a_D,a)$ with the harmonic functions $(F,G)$, and $\tau$ with the torus
modulus $\tau^3=\tau$.  Furthermore, the position $z=\pm L$ of the singularities on
the moduli space is identified with the position of the supertubes in
the near region.  The precise identification between $(F,G)$ and
$(a_D,a)$ is
\begin{align}
 \begin{pmatrix} F \\ G \end{pmatrix}
 =  c \begin{pmatrix} a_D'(z) \\ a'(z) \end{pmatrix}
\label{(F,G)_(aD',a')}
\end{align}
where $c\in\bbC$ is a constant of dimension
$[c]=(\text{length})^{1/2}$.\footnote{At this stage, $c$ can actually be
an arbitrary single-valued holomorphic function in $z$.  However, one
can show that, in order that the fields near each of the two supertube
at $z=\pm L$ behave the same way as they do near ordinary supertubes,
such as the $\text{D2}+\text{D2}\to \text{ns5}$ supertube given in
\eqref{FG_D2D2ns5} or the $\text{D2}+\text{D6}\to 5^2_2$ supertube given
in \eqref{FG_D2D6522}, we must take $c$ to be constant.  It must be
possible to derive the behavior of $c$ near supertubes by properly
taking account of its backreaction of the brane worldvolume. See
\cite{Bergshoeff:2006jj} for a discussion of such backreaction in
F-theory configurations of 7-branes.}  Now Figure~\ref{fig:2dplane} is understood as
the monodromy structure of the harmonic functions $(F,G)$ in the near
region.

One may wonder about the meaning, in the supertube context, of the
singularity at $z=\infty$ of the Seiberg-Witten solution.  Recall that
the near-region description in terms of the $z$-plane is only an
approximation near the tubes.  In reality, the infinity of the
near-region $z$-plane is connected to the 3D space, where the tube is
not infinitely long but is finite and closed.  In the context of the
original Seiberg-Witten theory, which is defined in the $z$-plane, the
monodromy at $z=\pm L$ must be canceled by the monodromy at $z=\infty$.
On the other hand, in the supertube context, the $z$-plane is connected
to a larger space, $\bbR^3$ and the monodromy is canceled by the other
side of the supertube in $\bbR^3$.

The explicit expression for $a(z)$ and $a_D(z)$ is
\begin{equation}
\begin{split}
a(z)
&=
\frac{\sqrt{2}}{\pi}\int_{-L}^{L}\ud{x}
 \sqrt{z-x \over (L-x)(L+x)}
=
\sqrt{2(z+L)}\, {}_2F_1\! \left(-\frac{1}{2},\frac{1}{2};1;\frac{2L}{z+L}\right)
\,,\\
a_D(z)
&=
\frac{\sqrt{2}\,i}{\pi}\int_L^{z}\ud{x}\sqrt{z-x\over (x-L)(x+L)}
=
\frac{L-z}{2i\sqrt{L}}\, {}_2F_1\!\left(\frac{1}{2},\frac{1}{2};2;\frac{L-z}{2L}\right)\,.
\end{split}
\label{a,aD_def}
\end{equation}
Here $_2F_1(a,b;c;z)$ is the hypergeometric function.  Note that $L$ is
a complex number (see~\eqref{L=|L|e^il}). The sign of the square root in
the integral expression is defined to be positive for $0<L<z$ and, for complex
$L,z$, it is defined by analytic continuation.
%
Taking derivatives, we have
\begin{align}
\begin{split}
 a'(z)&=
\frac{1}{\sqrt{2}\,\pi}\int_{-L}^{L}
 {\ud{x}\over \sqrt{(z-x)(L-x)(L+x)}}
 =
 \frac{\sqrt{2}}{\pi\sqrt{z+L}}\,K\!\left(\frac{2L}{z+L}\right)\,,
 \\
 a_D'(z)
 &=
 \frac{i}{\sqrt{2}\,\pi}\int_L^{z}{\ud{x}\over \sqrt{(z-x)(x-L)(x+L)}}
=\frac{i}{\pi\sqrt{L}}\,K\!\left(\frac{L-z}{2L}\right)\,,
\end{split}
\label{a',aD'_def}
\end{align}
where $K(z)=\frac{\pi}{2}\,{}_2F_1(\frac{1}{2},\frac{1}{2};1;z)$ is the
complete elliptic integral of the first kind.  As mentioned above, as we
go around the singular points $z=L,-L$ and $z=\infty$, the pair
$(a_D,a)$ and hence $(a_D',a')$ undergoes $\text{SL}(2,\bbZ)$
transformations given by the monodromy matrices $M_1,M_2$ in
\eqref{M1M2} and $M$ in \eqref{Mtotal}, respectively.

Now we have $(F,G)$ in the near region, which is related via
\eqref{(F,G)_(aD',a')} to $(a_D',a')$ given in~\eqref{a',aD'_def}.  To
match this with the far-region solution, we will later need the
$|z|\to\infty$ behavior of $(a_D',a')$.  It is given by
\begin{subequations}
\label{apadpinf}\\
\begin{align}
a'(z)&=\frac{1}{\sqrt{2z}}+\frac{3L^2}{4(2z)^{5/2}}+\frac{105L^4}{64(2z)^{9/2}}+\dotsb\,,\label{apzinf}\\
a_D'(z)&=\frac{i}{\pi}\left[\frac{1}{\sqrt{2z}}\ln\frac{8z}{L}+\frac{3L^2}{4(2z)^{5/2}}\left(\ln\frac{8z}{L}-\frac{5}{3}\right)+\frac{105L^4}{64(2z)^{9/2}}\left(\ln\frac{8z}{L}-\frac{389}{210}\right)+\dotsb\right]\label{adpzinf}\,.
\end{align}
\end{subequations}
Just from the leading terms, it is easy to check that we have the
monodromy
\begin{equation}
\begin{pmatrix}
a_D' \\ a'
\end{pmatrix}\to\begin{pmatrix}
-1 & 2 \\ 0 & -1
\end{pmatrix}\begin{pmatrix}
a_D' \\ a'
\end{pmatrix}=M\begin{pmatrix}
a_D' \\ a'
\end{pmatrix}\,.
\end{equation}

For later convenience, let us also write down the 
behavior near the singularities $z=\pm L$.  Near $z=L$,
\begin{subequations}
\label{apaDp_zL}
\begin{align}
a'(z)
&=
-\frac{1}{2\pi\sqrt{L}}\left[
\ln\frac{z-L}{32L}
-\frac{1}{8L}\left(\ln\frac{z-L}{32L}+2\right)(z-L)
+\dotsb\right].\\
a_D'(z)
&=
\frac{i}{2\sqrt{L}}\left[
1
-\frac{1}{8L}(z-L)
+\dotsb\right]=\frac{i}{2\sqrt{L}}\sum_{n=0}^\infty\left(\frac{(2n)!}{2^{2n}n!^2}\right)^2\left(\frac{-1}{2L}\right)^n(z-L)^n.
\end{align}
\end{subequations}
Near $z=-L$,
\begin{subequations}
\label{apaDp_zmL}
\begin{align}
a'(z)
&=
\frac{i}{2\pi\sqrt{L}}\left[
\ln\frac{z+L}{-32L}
+\frac{1}{8L}\left(\ln\frac{z+L}{-32L}+2\right)(z+L)
+\dotsb\right]. \\
a_D'(z)
&=
-\frac{i}{2\pi\sqrt{L}}\left[
\ln\frac{z+L}{32L}
+\frac{1}{8L}\left(\ln\frac{z+L}{32L}+2\right)(z+L)
+\dotsb\right].
\end{align}
\end{subequations}
From these, it is easy to check the monodromy $M_1,M_2$.


\subsection{The far region: coordinate system and boundary conditions}
\label{sec:coordbdy}

Having fixed the near-region solution, the next task is to find the
far-region solution that matches onto it.  For that, as preparation, let
us introduce the coordinate system appropriate for our purpose and
discuss the boundary conditions that the far-region solution must
satisfy.

\begin{figure}[h]
\capstart
\begin{quote}
 \begin{center}
 \includegraphics[width=0.35\textwidth]{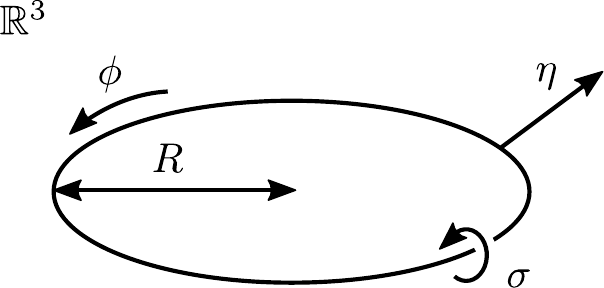}
 \caption{{\sl Toroidal coordinates $(\eta,\sigma,\phi)$. $\eta$ is a
 ``radial'' coordinate that decreases as one goes away from the ring,
 $\sigma$ is the angular variable around the ring and $\phi$ is an
 angular variable along the ring.}} \label{tor3d}
 \end{center}
\end{quote}
\end{figure}

\begin{figure}[h]
\begin{quote}
 \capstart
 \begin{center}
 \includegraphics[height=0.5\textwidth]{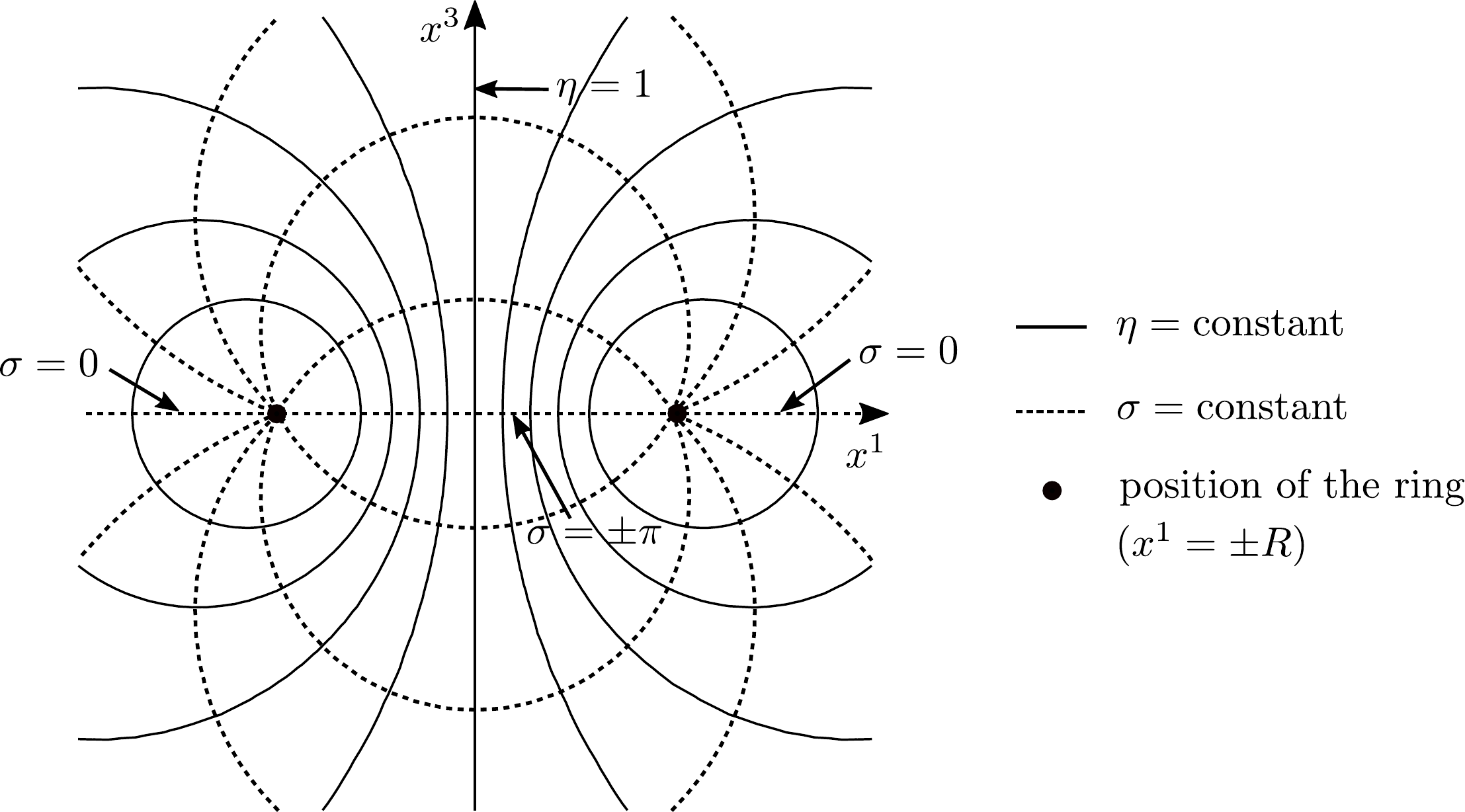}
 \caption{\sl Toroidal coordinates in the $x^2=0$ section. Solid lines
 represent constant-$\eta$ surfaces and dotted lines represent
 constant-$\sigma$ surfaces. As $\eta\to1$, the constant-$\eta$ surface
 approaches the vertical ($x^3$) axis , while the position of the ring
 corresponds to the $\eta\to\infty$ limit.}
\label{tor2d}
 \end{center}
\end{quote}
\end{figure}

\subsubsection*{Toroidal coordinate system}

As we explained, in the far region, we effectively have one
supertube. To describe this configuration, we introduce the toroidal
coordinate system $(\eta,\sigma,\phi)$
\cite{Morse:1953mtp}; see Figures~\ref{tor3d}
and~\ref{tor2d}. In terms of Cartesian coordinates $(x^1,x^2,x^3)$, the
toroidal coordinates are given by
\begin{equation}
x^1=R\frac{\sqrt{\eta^2-1}}{\eta-\cos\sigma}\cos\phi\,,\quad x^2=R\frac{\sqrt{\eta^2-1}}{\eta-\cos\sigma}\sin\phi\,,\quad x^3=R\frac{\sin\sigma}{\eta-\cos\sigma}\,,\label{cartesian_to_toroidal}
\end{equation}
where $R$ is the radius of the ring, $\sigma$ is the angular variable around the ring and $\phi$ is the angular variable along the ring. The inverse relations are given by
\begin{equation}
\eta=\frac{\vect{x}^2+R^2}{\Sigma}\,,\quad \cos\sigma=\frac{\vect{x}^2-R^2}{\Sigma}\,,\quad \tan\phi=\frac{x^2}{x^1}\,,\label{toroidal_ito_cartesian}
\end{equation}
with
\begin{equation}
\Sigma^2=(\vect{x}^2-R^2)^2+4R^2(x^3)^2\,.\label{Sigma_def}
\end{equation}
The domain of the coordinates is $1\le\eta<\infty$, $-\pi\le\sigma<\pi$,
$0\le\phi<2\pi$. Then, the flat 3D metric in the toroidal coordinates is
given by
\begin{equation}
ds^2=\frac{R^2}{(\eta-\cos\sigma)^2}\left(\frac{\ud{\eta}^2}{\eta^2-1}+\ud{\sigma}^2+(\eta^2-1)\ud{\phi}^2\right)\,.
 \label{3Dmetric_toroidal}
\end{equation}

To connect the far- and near-region solutions, we have to relate the
near-region (2D) and the far-region (3D) coordinates. In the near-region
limit $\eta\to\infty$, the Cartesian coordinates are given, to leading
order, by
\begin{equation}
x^1\simeq R+\frac{R\cos\sigma}{\eta}\,,\qquad
x^2=0\,,\qquad
x^3\simeq \frac{R\sin\sigma}{\eta}\,.
\end{equation}
Then we can relate the $z$ coordinate defined in \eqref{z_and_R3} to the toroidal coordinates $(\eta,\sigma)$ as
\begin{equation}\label{ztoeta}
z=(x^1-R)+ix^3=\frac{R}{\eta}e^{i\sigma}\,.
\end{equation}
This is the fundamental relation to connect the near- and far-region
solutions.

\subsubsection*{Boundary conditions}

On the far-region solution, we have to impose boundary conditions at
infinity ($\eta\to1$ and $\sigma\to0$ simultaneously) and near the
supertube ($\eta\to\infty$).

First, let us discuss the boundary condition at infinity. We require
the harmonic functions to go as
\begin{equation}\label{bdcinf}
H=h+ \frac{\Gamma}{r}+\cO\left({1\over r^2}\right)
 \quad\text{as}\quad r\to\infty,
\end{equation}
where $r=\sqrt{(x^1)^2+(x^2)^2+(x^3)^2}$. This is the same $r\to\infty$
behavior as the codimension-3 solution, \eqref{codim3harm} (or
\eqref{harm_FG}).  This is because we are interested in codimension-2
branes (supertubes) which have been produced by the supertube transition
out of codimension-3 branes.  Very far from it, the codimension-2 brane
must look like a codimension-3 object with the original monopole charge.
Therefore, the harmonic function must have the $1/r$ term whose
coefficient $\Gamma$ is the same as the total monopole charge of the
original brane configuration.

The boundary condition near the tube $(\eta\to\infty)$ comes from the
matching condition discussed at the end of
Section \ref{sec:strategy}. Let us write the large-$|z|$ expansion of
$a'(z)$ and $a_D'(z)$ as\footnote{This expansion corresponds to \eqref{kahc20Apr17} of the toy model in Section \ref{sec:strategy}.}
\begin{equation}\label{asympexpaad}
a'(z)=\sum_{n=0}^\infty a'_{n}(z)
\,,\qquad
a_D'(z)=\sum_{n=0}^\infty a'_{D n}(z)
\,,
\end{equation}
where $a'_n,a'_{D n}=\cO(z^{-2n-1/2})$ (here it is understood that
$\cO(z^{-2n-1/2})$ includes $z^{-2n-1/2}\log z$).  The first three terms of
each expansion are given in \eqref{apzinf} and \eqref{adpzinf}. As we
discussed earlier in Section~\ref{sec:strategy}, we must be able to find
a far-region solution that matches onto this expansion, order by
order. Concretely, let us do a near-ring ($\eta\to \infty$) expansion of
the far-region harmonic functions $F$ and $G$ and let the $n$-th term be
$F_n$ and $G_n$ where their behavior as $\eta\to\infty$\footnote{The behavior will be determined in the next section \ref{sec:far} and Appendix \ref{app:perturb}.} is
$F_n,G_n=\cO(\eta^{2n+1/2})$.\footnote{These $n$-th terms correspond to
\eqref{mepq19Aug17} of the toy model in Section \ref{sec:strategy}.}
Then, upon using the dictionary \eqref{ztoeta}, we must have
\begin{equation}
\label{bdctube}
F_n = c a'_{D n}+\cO(\eta^{2n-1/2}),\qquad
G_n = c a'_{n}+\cO(\eta^{2n-1/2}),\qquad
\eta\to \infty.
\end{equation}
Note that the lesson of the toy model in Section \ref{sec:strategy} was
that we have to take the limit $r\ll R$ first, and then match the small
$\frac{\abs{L}}{r}$ expansion.  In the present case, the former 
corresponds to matching only the leading $\cO(\eta^{2n+1/2})$ term
in \eqref{bdctube}, while the latter corresponds to
doing this for each value of $n$.

For example, for the first ($n=0$) term, we have
\begin{equation}\label{bdctube1}
F_0=\frac{ic}{\pi\sqrt{2z}}\ln\frac{8z}{L}+\cO(\eta^{-1/2})
\,,\qquad
G_0=\frac{c}{\sqrt{2z}}+\cO(\eta^{-1/2})
\,.
\end{equation}
In principle, we can find $F_n$ and $G_n$ satisfying \eqref{bdctube} for $n$
arbitrarily large.  If we could carry out the infinite sum $F=\sum_n
F_n$ and $G=\sum_n G_n$, it would correspond to the exact two-supertube
solution defined in the entire $\bbR^3$.


\subsection{The far region: the solution}
\label{sec:far}


In the far region, there is only one supertube (see Figure~\ref{tor3d}) and
we are instructed to find a pair of harmonic functions $(F,G)$ that 
has the monodromy
\begin{equation}
\begin{pmatrix}F \\ G\end{pmatrix}
\to
M
\begin{pmatrix}F \\ G\end{pmatrix}
=
\begin{pmatrix}-1 & 2 \\ 0 & -1\end{pmatrix}
\begin{pmatrix}F \\ G\end{pmatrix}\,
\end{equation}
as $\sigma\to\sigma+2\pi$. In other words,
\begin{subequations} 
 \begin{align}
 F&\to -F+2G\,,\label{montype1}\\
 G&\to -G\,,\label{montype2}
 \end{align}
\end{subequations}


\subsubsection*{Harmonic functions in toroidal coordinates}

Let us explain now how to construct $F$ and $G$. We start with the
ansatz for $G$ since its monodromy \eqref{montype2} is simpler. If we assume
the following separated form,
\begin{equation}\label{harmult}
G(\eta,\sigma,\phi)=\sqrt{\eta-\cos\sigma}\,T(\eta)S(\sigma)V(\phi)\,,
\end{equation}
the Laplace equation becomes
\begin{align}
\Delta G&=\frac{(\eta-\cos\sigma)^{5/2}}{R^2}T(\eta)S(\sigma)V(\phi)\notag\\
&\quad\times\left[\frac{1}{\eta^2-1}\frac{V''(\phi)}{V(\phi)}+\frac{S''(\sigma)}{S(\sigma)}+\frac{1}{T(\eta)}\left(\frac{1}{4}T(\eta)+2\eta T'(\eta)+(\eta^2-1)T''(\eta)\right)\right]\notag\\
&=0\,.\label{signansatz}
\end{align}
This can be reduced to the following three ordinary differential
equations:
\begin{subequations}
\begin{align}
0&=V''(\phi)+m^2V(\phi)\,,\label{odeVphi}\\
0&=S''(\sigma)+k^2S(\sigma)\,,\label{odeSsigma}\\
0&=(\eta^2-1)T''(\eta)+2\eta T'(\eta)+\left(\frac{1}{4}-k^2-\frac{m^2}{\eta^2-1}\right)T(\eta)\,,\label{odeTeta}
\end{align}
\end{subequations}
with arbitrary constants $m$ and $k$. The general solutions for
these equations are given by
\begin{subequations}
\label{harVphiSsigmaTeta}
\begin{align}
V(\phi)&=e^{im\phi}\,,\label{harVphi}\\
S(\sigma)&=e^{ik\sigma}\,,\label{harSsigma}\\
T(\eta)&=P_{\abs{k}-1/2}^{\abs{m}}(\eta)\quad\text{and}\quad Q_{\abs{k}-1/2}^{\abs{m}}(\eta)\,,\label{harTeta}
\end{align}
\end{subequations}
where $P_k^m(\eta)$ and $Q_k^m(\eta)$ are the associated Legendre
functions of the first and second kind, respectively, with degree $k$
and order $m$.  If we require $2\pi$ periodicity along the $\phi$
(respectively $\sigma$) direction, the constant $m$ (respectively $k$) will take
integer values. Because our configuration is symmetric along $\phi$ (see
Figure \ref{tor3d}), we should take $m=0$. Then as we can easily see
from the form of the solutions \eqref{harVphiSsigmaTeta}, we have to
choose $k\in\mathbb{Z}+1/2$ in order for $G$ to have the
monodromy \eqref{montype2}. So the solution for $G$ is written as
\begin{equation}\label{presolG}
G=\sqrt{\eta-\cos\sigma}\,e^{ik\sigma}\left(A_{\abs{k}-1/2}P_{\abs{k}-1/2}(\eta)+B_{\abs{k}-1/2}Q_{\abs{k}-1/2}(\eta)\right)\,,
\end{equation}
where $k\in\mathbb{Z}+1/2$ and $A_{\abs{k}-1/2},B_{\abs{k}-1/2}$ are constants.

Let us turn to $F$. The monodromy \eqref{montype1} motivates
the  following ansatz:
\begin{equation}\label{haradd}
F(\eta,\sigma,\phi)=\sqrt{\eta-\cos\sigma}\left(U(\eta)-\frac{\sigma}{\pi}\,T(\eta)\right)S(\sigma)V(\phi)\,.
\end{equation}
Plugging this into the Laplace equation, we obtain
\begin{align}
0&=U(\eta)\left[\frac{1}{\eta^2-1}\frac{V''(\phi)}{V(\phi)}+\frac{S''(\sigma)}{S(\sigma)}+\frac{1}{U(\eta)}\left(\frac{1}{4}U(\eta)+2\eta U'(\eta)+(\eta^2-1)U''(\eta)\right)-\frac{2}{\pi}\frac{T(\eta)}{U(\eta)}\frac{S'(\sigma)}{S(\sigma)}\right]\notag\\
&\quad-\frac{\sigma}{\pi}\, T(\eta)\left[\frac{1}{\eta^2-1}\frac{V''(\phi)}{V(\phi)}+\frac{S''(\sigma)}{S(\sigma)}+\frac{1}{T(\eta)}\left(\frac{1}{4}T(\eta)+2\eta T'(\eta)+(\eta^2-1)T''(\eta)\right)\right]
\,.\label{addansatz}
\end{align}
If we take $T,S$ and $V$ to be the solutions of \eqref{signansatz} given by
\eqref{harVphiSsigmaTeta}, then
the second line of \eqref{addansatz} vanishes and we are left with
\begin{equation}
\label{odeUeta}
(\eta^2-1)U''(\eta)+2\eta U'(\eta)+\left(\frac{1}{4}-k^2-\frac{m^2}{\eta^2-1}\right)U(\eta)=\frac{2}{\pi}\,T(\eta)\frac{S'(\sigma)}{S(\sigma)}\,.
\end{equation}
This differential equation differs from \eqref{odeTeta} in its
inhomogeneous term. The solution of \eqref{odeUeta} for a specific
choice of $T(\eta)$ and $S(\sigma)$ can be easily found. We gave a few
examples in Appendix \ref{app:perturb}.

Even though we have to solve \eqref{odeUeta} to get explicit harmonic
functions, the monodromy can be easily seen without solving it. Let us
assume $k\in\mathbb{Z}+1/2$ as in \eqref{presolG} to get an overall
sign flip after going around the supertube $(\sigma\to\sigma+2\pi)$. We
also set $m=0$ because of the symmetry of our configuration. Then the
monodromy is exactly what we want \eqref{montype1}:
\begin{equation}
F\to-F+2G\quad\text{as}\quad\sigma\to\sigma+2\pi\,.
\end{equation}


If we choose a particular term in \eqref{asympexpaad} with a specific
value of $n$ that we want to reproduce, the value of $k$ can be
determined and the equation \eqref{odeUeta} can be solved. Here we
will focus on the first ($n=0$) term in \eqref{bdctube}. The leading
term in the large-$|z|$ expansion of $a'(z)$ is
\begin{equation}
a'_0 = \frac{1}{\sqrt{2z}}=\sqrt{\frac{\eta}{2R}}e^{-i\sigma/2}\,,
\end{equation}
where we have used the dictionary \eqref{ztoeta}. Then we have to take
$k=-1/2$ to reproduce this as a limit of the 3D harmonic function
$G$. We can easily show that this is also correct choice for $a_{D0}'$ and $F$.
With this choice, $T(\eta)$ is also fixed and is given by a linear
combination of $P_0(\eta)$ and $Q_0(\eta)$.

The resulting harmonic functions can be written as
\begin{align}
F(\eta,\sigma,\phi)
&=\sqrt{\eta-\cos\sigma}\,e^{-i\sigma/2}U(\eta)-\frac{\sigma}{\pi}G
\,,\\
G(\eta,\sigma,\phi)
&=\sqrt{\eta-\cos\sigma}\,e^{-i\sigma/2}T(\eta)
\,,
\end{align}
where
\begin{align}
 T(\eta)=A_0P_0(\eta)+B_0Q_0(\eta)
\end{align}
and
$U(\eta)$ is a solution of
\begin{equation}\label{odeUeta1}
(\eta^2-1)U''(\eta)+2\eta U'(\eta)=-\frac{i}{\pi}\,T(\eta)\,.
\end{equation}
$A_0$ and $B_0$ are constant of integration which should be chosen from
the boundary conditions.

It is easy to write down solutions explicitly if we impose boundary
conditions at infinity, \eqref{bdcinf}, before solving
\eqref{odeUeta1}. The boundary condition at infinity, \eqref{bdcinf},
leads to the condition
\begin{equation}
B_0=0
\,,
\end{equation}
since $Q_0(\eta)$ diverges at 3D
infinity.\footnote{\label{ftnt:Q_div} More precisely, $B_0\neq 0$ would
lead to divergence at 3D infinity and on the $x^3$-axis.  If
$\sigma\neq 0$, as we can see from \eqref{cartesian_to_toroidal},
$\eta=1$ corresponds to the points on the $x^3$-axis,
$(x^1,x^2,x^3)=(0,0,{R\cot{\sigma\over 2}})$.  As $\eta\to 1$,
$Q_{|k|-1/2}$ diverges as $\log(\eta-1)$ while the prefactor is
finite: $\sqrt{\eta-\cos\sigma}=\sqrt{2}\,|\!\sin{\sigma\over
2}|$. Therefore, $B_0\neq 0$ makes the harmonic function diverge on the
$x^3$-axis and should be avoided.} Then \eqref{odeUeta1} is easily
solved to give
\begin{equation}
U(\eta)=C_0P_0(\eta)+D_0Q_0(\eta)-\frac{i}{\pi}A_0\ln\frac{\eta+1}{2}
\,.
\end{equation}
By imposing the same boundary condition at infinity on $U(\eta)$, \eqref{bdcinf}, we conclude that
\begin{equation}
D_0=0
\,.
\end{equation}

The final expression for the harmonic functions is
\begin{align}
F(\eta,\sigma,\phi)
&=\sqrt{\eta-\cos\sigma}\,e^{-i\sigma/2}\frac{i}{\pi}A_0\left(\frac{\pi}{i}\frac{C_0}{A_0}-\ln\frac{\eta+1}{2}+i\sigma\right)
\,,\\
G(\eta,\sigma,\phi)
&=\sqrt{\eta-\cos\sigma}\,e^{-i\sigma/2}A_0
\,,
\end{align}
where we used $P_0(\eta)=1$.


\subsubsection*{Matching}

We have obtained the solutions in the near and far regions.  Let us fix
the coefficients $A_0$ and $C_0$ by matching the two solutions in the
intermediate region. This amounts to imposing the conditions
\eqref{bdctube1}. The near-ring ($\eta\to\infty$) expressions for $F$
and $G$ are
\begin{align}
F\simeq \sqrt{\eta}\,e^{-i\sigma/2}\frac{i}{\pi}A_0\left(\frac{\pi}{i}\frac{C_0}{A_0}-\ln\frac{\eta}{2}+i\sigma\right)
\,,\qquad
G\simeq \sqrt{\eta}\,e^{-i\sigma/2}A_0
\,.
\end{align}
Therefore, the conditions \eqref{bdctube1} read
\begin{equation}\label{matching1order}
\begin{split}
\frac{i}{\pi}\sqrt{\eta}\,e^{-i\sigma/2}A_0\left(\frac{\pi}{i}\frac{C_0}{A_0}-\ln\frac{\eta}{2}+i\sigma\right)
&=
\frac{i}{\pi}c\sqrt{\frac{\eta}{2R}}\,e^{-i\sigma/2}\left(\ln\frac{4R}{L}-\ln\frac{\eta}{2}+i\sigma\right)
\,,\\
\sqrt{\eta}\,e^{-i\sigma/2}A_0
&=
c\sqrt{\frac{\eta}{2R}}\,e^{-i\sigma/2}
\,.
\end{split}
\end{equation}
These determine the constants to be
\begin{align}\label{0thcoeff}
A_0=\frac{c}{\sqrt{2R}}\,,\qquad C_0=\frac{i}{\pi}\frac{c}{\sqrt{2R}}\ln\frac{4R}{L}\,.
\end{align}
The final expression for the far-region solution is
\begin{subequations}
\label{1ordersol}
\begin{align}
F(\eta,\sigma,\phi)&=\frac{ic}{\pi\sqrt{2R}}\sqrt{\eta-\cos\sigma}\,e^{-i\sigma/2}\left[-\ln\frac{L(\eta+1)}{8R}+i\sigma\right]\,,\\
G(\eta,\sigma,\phi)&=\frac{c}{\sqrt{2R}}\sqrt{\eta-\cos\sigma}\,e^{-i\sigma/2}\,.
\end{align}
\end{subequations}



\section{Physical properties of the solution}
\label{sec:physprop}


In the previous section, we obtained the explicit expression for the
harmonic functions $(F,G)$ in \eqref{1ordersol} which describes the
far-region behavior of a non-Abelian two-supertube configuration, at the
leading order in a perturbative expansion.  In terms of these complex
harmonic functions, the real harmonic functions $\{V,K^I,L_I,M\}$ can be
expressed via \eqref{tau12harm}.  Here we discuss some physical
properties of this solution.

\subsection{Geometry and charges}
\label{ss:geom_chg}

First, let us study the asymptotic form of the harmonic functions near
3D infinity, $r=\infty$, which corresponds to $\eta=1,\sigma=0$ in
the toroidal coordinates. Using the
relation~\eqref{toroidal_ito_cartesian}, we find that 
\begin{align}
F&=h_F+{Q_F\over r}+\cO\left({1\over r^2}\right)\,,\qquad
G =h_G+{Q_G\over r}+\cO\left({1\over r^2}\right)\,,
\label{FG_asympt}
\end{align}
where
\begin{align}
h_F&=h_G=0,\label{hFhG=0}
\\
Q_F&= i c\sqrt{R}\,\nu,\qquad
Q_G= c\sqrt{R}
\label{QFQG_def}
\end{align}
with
\begin{align}
\nu\equiv{1\over \pi}\log {4R\over L}.
\label{nu_def}
\end{align}
The asymptotic form \eqref{FG_asympt} is the same as that of the general
codimension-3 harmonic function,~\eqref{harm_FG}.  Note that, under our
assumption \eqref{colliding_lim},
\begin{align}
 \Re \nu = {1\over \pi}\log{4R\over |L|}\gg 1.\label{Renu>>1}
\end{align}

The asymptotic monopole charges of the solution can be read off from the
coefficients of the $1/r$ terms in the harmonic functions,
\eqref{QFQG_def}.  The corresponding D-brane numbers $N^0,N^I,N_I,N_0$
can be determined from the relation \eqref{QFQG_q'n}.  Explicitly,
\begin{align}
\label{asymptN_tube}
\begin{split}
  N^3+iN_1&={2i c\sqrt{R}\,\nu\over g_s l_s},\qquad
 N^0-iN^1={2c\sqrt{R}\over g_s l_s}
.
\end{split}
\end{align}
The entropy of the single-center black hole with charges \eqref{QFQG_def}
can be computed using \eqref{S_QFQG}:
\begin{align}
 S={8\pi\,|\!\Im(Q_F \Qb_G)|\over g_s^2 l_s^2}
 ={8\pi |c|^2 R \over g_s^2 l_s^2}\Re \nu.
\end{align}
This is non-vanishing because of \eqref{Renu>>1} and therefore our
solution has the same asymptotic charges as a black hole with a finite
horizon area.

One peculiar thing about the harmonic functions \eqref{FG_asympt} is
that the constant terms always vanish, $h_F=h_G=0$.  This fact came from
the harmonic analysis in the toroidal coordinates.  For example, in the
ansatz for $G$, \eqref{presolG}, the prefactor goes as
$\sqrt{\eta-\cos\sigma}\sim \sqrt{2}R/r$ in the 3D infinity limit
$\eta\to 1,\sigma\to 0$. On the other hand, $P_{|k|-1/2}(\eta=1)=1$
and therefore $G\sim 1/r$ and does not have a constant term.  We do not
have the option of turning on $Q_{|k|-1/2}(\eta)$, because it diverges
on the $x^3$-axis and should not be present (see Footnote
\ref{ftnt:Q_div}).

This means that this solution cannot have flat asymptotics.  Instead,
the asymptotic geometry is always the attractor geometry
\cite{Ferrara:1995ih} of a single-center black hole with D6, D4, D2 and
D0 charges in the near-horizon limit.
Indeed, the asymptotic form of the type IIA geometry is
easily seen from \eqref{IIAfield} to be
\begin{subequations} 
\label{IIAfields_FG}
 \begin{align}
 ds_{10,\text{str}}^2
 &=
 -{1\over \Im(F\Gb)}(dt+\omega)^2+ \Im(F\Gb)\left(dr^2+r^2d\Omega_2^2\right)
 +dx_{4567}^2+\Im\left({F\over G}\right)dx_{89}^2
 \notag\\
 &\sim
 -{r^2\over \Im(Q_F\Qb_G)}dt^2+ \Im(Q_F\Qb_G)\left({dr^2\over r^2}+d\Omega_2^2\right)
 +dx_{4567}^2+\Im\left({Q_F\over Q_G}\right)dx_{89}^2,\\
 e^{2\Phi}&=\Im\left({F\over G}\right)\sim \Im\left({Q_F\over Q_G}\right).
 \end{align}
\end{subequations}
We see that this is AdS$_2\times S^2\times T^6$ with radius $\cR_{\rm
AdS_2}=\cR_{S^2}=\sqrt{\Im(Q_F\Qb_G)}$.

\subsubsection*{Asymptotic charge versus local charge}

It is interesting to compare the asymptotic charges \eqref{QFQG_def}
with the one that we would obtain from the behavior of fields near the
supertubes.
From \eqref{apaDp_zL} and \eqref{apaDp_zmL}, we find that the behavior
of the harmonic functions $F,G$ near the supertubes is
\begin{align}
\begin{aligned}
 z\sim +L&:\quad & F&\sim \text{const.},&  G&\sim -{c\over 2\pi\sqrt{L}}\log(z-L),\\
 z\sim -L&:& F&\sim -{ic\over 2\pi\sqrt{L}}\log(z+L),&
 G&\sim {ic\over 2\pi\sqrt{L}}\log(z+L).
\end{aligned}
\label{FGneartube_z}
\end{align}
If a codimension-2 source at $|z|=0$ has D-brane number densities $n^0$,
$n^1$, $n^3$ and $n_1$ per unit length for D6(456789), D4(6789), D4(4567), and
D2(45) branes, respectively, then the harmonic functions will have the
following logarithmic behavior:\footnote{For example, if we array
D6-branes at intervals of distance $a$, from \eqref{charge_q'n}
\begin{align}
 V\sim {g_s l_s \over 2}\sum_{n\in\bbZ} {1\over\sqrt{|z|^2+na}}
 \sim {g_s l_s \over 2a}\int_{-\Lambda}^{\Lambda}  {dx\over\sqrt{|z|^2+x^2}}
 \sim -{g_s l_s \over a}\log {|z|\over 2\Lambda}+\cO(\Lambda^{-2})
\end{align}
where $\Lambda$ is a cutoff. By replacing $a$ with $1/n^0$, we obtain
\eqref{VKLM_n}.
}
\begin{align}
\begin{split}
  V&\sim -g_s l_s n^0 \log |z|,\qquad
 K^1\sim -g_s l_s n^1 \log |z|,\\
 K^3&\sim -g_s l_s n^3 \log |z|,\qquad
 L_1\sim -g_s l_s n_1\log |z|.
\end{split}
\label{VKLM_n}
\end{align}
Or, in terms of the complex harmonic functions $F,G$,
\begin{align}
F
\sim -g_s l_s (n^3+in_1)\log|z|,\qquad
G
 \sim -g_s l_s (n^0-in^1)\log|z|.
\end{align}
Comparing this with \eqref{FGneartube_z}, we see that the D-brane number
densities are
\begin{align}
\begin{aligned}
 z=+L&:&\qquad
 n^3+i n_1 &=0,&  n^0-i n^1 &= {c\over 2\pi g_s l_s \sqrt{L}},
 \\
 z=-L&:& 
 n^3+i n_1 &={ic\over 2\pi g_s l_s \sqrt{L}},&  n^0-i n^1 &= -{ic\over 2\pi g_s l_s \sqrt{L}}.
\end{aligned}
\end{align}
Because these charges are distributed over rings of radius approximately
$R$, the total D-brane numbers would be
\begin{align}
 N^3+i N_1 &\stackrel{?}{=}{ic R\over  g_s l_s \sqrt{L}},\qquad
 N^0-i N^1 \stackrel{?}{=} {(1-i)cR \over  g_s l_s \sqrt{L}}.
\label{localN_tube}
\end{align}
These are completely different from the charge we observe at infinity,
\eqref{asymptN_tube}.

The reason why we obtained incorrect total charges \eqref{localN_tube}
is that our solution is multi-valued.  In normal situations, the
Gaussian surface on which we integrate fluxes to obtain charges can be
continuously deformed from asymptotic infinity to small surfaces
enclosing local charges.  However, in the present case, the fields in
our solution are multi-valued because of the monodromies around the
supertubes, and so are the fluxes.  Another way of saying this is that
there is a branch cut (or disk) inside each of the two tubes, and the
fluxes are discontinuous across it.  When we deform the Gaussian surface
at infinity, we cannot shrink them to enclose just the supertubes; all
we can do is to deform it into two surfaces, each of which encloses one
entire branch disk with the supertube on its circumference.  When we
evaluate the flux integral on the Gaussian surfaces, there will be
contributions not just from the supertubes but also from (the
discontinuity in) the fluxes on the disks.  The difference between
\eqref{asymptN_tube} and \eqref{localN_tube} is due to the contribution
from the fluxes on the disks.

This situation of branch cuts carrying charge by the discontinuity in the
fluxes across it is an example of the so-called Cheshire charge that
appears in the presence of vortices with non-trivial monodromies called
Alice strings \cite{Schwarz:1982ec,Alford:1990mk,Preskill:1990bm}. For
discussions on the realizations of Alice strings in string theory, see
\cite{Harvey:2007ab, Okada:2014wma}.

When integrating fluxes on Gaussian surfaces to compute charges in the
presence of Chern-Simons interactions (such as supergravity in 11, 10,
and 5 dimensions), one must be careful about different definitions
of charges \cite{Marolf:2000cb}.  The relevant one here is the Page charge,
which is conserved, localized, quantized, and gauge-invariant under
small gauge transformations. For Page charge, we can freely deform a
Gaussian surface unless they cross a charge source or a branch cut for
the fluxes.  The discussion of charges in the paragraphs above is
understood to be using the Page charge.  For the explicit form of the
Page fluxes for D-brane charges, see, e.g.,~\cite[App.~D]{deBoer:2012ma}\cite[App.~E]{Park:2015gka}.

\subsubsection*{Angular momentum}

By solving equation \eqref{domega_FG}
for the harmonic functions given in \eqref{1ordersol}, we find
\begin{equation}\label{1orderomega}
\omega=\frac{\abs{c}^2}{2\pi}\left(\eta+1)\ln\frac{\abs{L}(\eta+1)}{8R}+2\ln\frac{4R}{\abs{L}}\right)\ud{\phi}\,,
\end{equation}
where the integration constant was fixed by requiring that $\omega$
vanish at $\eta=1$ (3D infinity).  In spherical polar coordinates
($r,\theta,\varphi$), the asymptotic behavior of \eqref{1orderomega} as
$r\to \infty$ is
\begin{equation}
\omega\simeq\frac{\abs{c}^2R^2}{\pi}\left(1+\ln\frac{\abs{L}}{4R}\right)\frac{\sin^2\theta}{r^2}\ud{\varphi}
 =\cO\!\left({1\over r^2}\right)\,.
\label{omega=O(1/r^2)}
\end{equation}
In four dimensions, the angular momentum is given by the
$\cO\!\left({1\over r}\right)$ term in the $(t,i)$ components of the
metric, which is nothing but the 1-form $\omega$ in our case. Therefore, we
conclude that the 4D angular momentum $J$ of our configuration vanishes:
\begin{align}
 J=0.
\end{align}
Note that \eqref{omega=O(1/r^2)} means that the entire angular momentum
vector vanishes, not just its $x^3$ component.

\subsection{Closed timelike curves}

No-CTC conditions for the one-modulus class solutions with $\tau^1=\tau^2=i$ were
briefly discussed in Section \ref{sec:taucases}.  For the explicit
harmonic functions of the far-region solution \eqref{1ordersol}, the condition \eqref{tau12CTCs}
gives
\begin{equation}\label{ctc1order}
\Im(F\bar{G})
 =\frac{\abs{c}^2(\eta-\cos\sigma)}{2\pi R}\ln\frac{8R}{\abs{L}(\eta+1)}
 \simeq \frac{\abs{c}^2 \eta}{2\pi R}\ln\frac{8R}{\abs{L}\eta}
\ge0\,
\end{equation}
for large $\eta$ (near the supertube).
This means that, in order not to have CTCs, we must restrict the range
of the variable $\eta$ to be
\begin{equation}
 \label{condeta}
\eta\lesssim\frac{8R}{\abs{L}}\,.
\end{equation}
Namely, the far-region solution has CTCs very near the tube.

Next, let us consider the positivity of the metric \eqref{omegacondCTCs}
along the supertube direction, $\phi$.  This gives
\begin{equation}\label{ctcmetric}
-\frac{\omega^2}{\mathcal{Q}}+\frac{R^2(\eta^2-1)}{(\eta-\cos\sigma)^2}\ud{\phi}^2\ge 0.
\end{equation}
After plugging the explicit expression for $\omega$
\eqref{1orderomega}, we can rewrite \eqref{ctcmetric} as
\begin{align}
& \frac{R^2\ud{\phi}^2}{(\eta-\cos\sigma)^2\left[\ln\frac{\abs{L}(\eta+1)}{8R}\right]^2}\notag\\
&\qquad\times\left((\eta^2-1)\left[\ln\frac{\abs{L}(\eta+1)}{8R}\right]^2-\left[(\eta+1)\ln\frac{\abs{L}(\eta+1)}{8R}+2\ln\frac{4R}{\abs{L}}\right]^2\right)\ge0\,.\label{ctcmetric2}
\end{align}
Near the ring ($\eta\to\infty$), the no-CTC condition \eqref{ctcmetric2} gives
\begin{equation}
-2\eta\,\ln\!\left(\frac{2R\eta}{\abs{L}}\right)\ln\!\left(\frac{\abs{L}\eta}{8R}\right)\ge0\,,
\end{equation}
which is satisfied for
\begin{equation}
\frac{\abs{L}}{2R}<1\le\eta\le\frac{8R}{\abs{L}}\,.
\end{equation}
The lower bound does not impose any condition on $\eta$ because
$\eta\ge 1$ by definition, and the upper bound is the same as
\eqref{condeta}.

So, we found that there are CTCs in the far-region solution very near
the ring, $\eta\sim {8R\over |L|}$.  However, this does \emph{not}
represent a problem with our solution.  It only indicates that, too
much near the ring, the description in terms of the far-region solution
with a single ring breaks down and we must instead switch to the
near-region solution with two rings. Indeed, by the relation
\eqref{ztoeta}, $\eta\sim {R\over |L|}$ corresponds to $|z|\sim |L|$ in
the near region, which is the distance scale at which the single
``effective'' supertube must be resolved into two supertubes.
This is exactly parallel to the familiar story in the context of
F-theory \cite{Sen:1996vd, Johnson:2003gi}.  In type IIB perturbative
string theory, the O7-plane has negative tension and its backreacted
metric has a wrong signature very near its worldvolume.  However, in
F-theory, non-perturbative effects resolve the O7-plane into two $(p,q)$
7-branes and replace the wrong-signature metric by a new metric with the
correct signature everywhere.  The two $(p,q)$ 7-branes have
non-commuting monodromies of the $\text{SL}(2,\bbZ)$ duality of type IIB
string.  We are seeing exactly the same phenomenon in a more
involved situation with circular supertubes.

To rigirously show that our solution is completely free from CTCs, we
must construct the exact solution by summing up the infinite
perturbative series, because the perturbative solution to any finite
order will have CTCs (this is related to the limitation of the matching
expansion discussed below \eqref{mepq19Aug17}).  However, that is beyond
the scope of the present paper and we will leave it as future research.

\subsection{Bound or unbound?}

Our 2-supertube configuration has three parameters: $c\in\bbC$ determines
the overall amplitude of the harmonic functions, $L\in\bbC$ parametrizes
the distance and the angle between two supertubes, and $R>0$ is the
average radius of the two supertubes.  The crucial question is: does
this represent a bound state or not?

In the case of codimension-3 solutions, allowed multi-center
configurations are determined by imposing equation
\eqref{balancingcond}. How this works is as follows.  One first fixes
the value of moduli (the constant terms in $H$), the number of centers
(say $N$), and the charges of each center ($\Gamma^p$, $p=1,\dots,N$).
By plugging these data into \eqref{balancingcond}, we can fix the
inter-center distances $a_{pq}$.  After this, some
parameters will remain unfixed. They parametrize the internal degrees of
freedom of the multi-center configuration, similar to the internal atomic
motion inside a molecule.  When it is a bound state, it is not
possible to take some centers infinitely far away from the rest of the
centers by tuning the parameters.

In our solution, the asymptotic moduli have already been fixed to the
attractor value~\cite{Ferrara:1995ih}. We have two codimension-2
supertube centers, and we know that the total monopole charges are given
by $(Q_F,Q_G)$. Actually, as we will discuss below, the monopole charges
of each of the two supertubes can be also determined if we fix the
complex charges $Q_F,Q_G$.  So, the question is whether there is some free
parameter left by tuning which we can make the two tubes infinitely far
apart. If so, then the configuration is unbound. Otherwise, it is bound.



Our solution contains five real parameters ($R\in\mathbb{R}$;
$c,L\in\mathbb{C}$) and four of them can be determined by fixing
$Q_{F,G}\in\bbC$. So, we seem to be left with one free real parameter.
For example, we can take it to be $|L|$, the absolute value of the
inter-tube distance parameter $L$.  If $|L|$ could take an arbitrarily
large value, the two tubes could be separated infinitely far away from
each other and thus the solution would be unbound.  Physically, however,
we expect that we can constrain this parameter by requiring the absence
of CTCs \cite{Emparan:2001ux, Mateos:2001pi}, and that the tubes cannot
be infinitely separated.  Such no-CTC analysis would be possible
\emph{if we knew the exact solution}. The problem is that we
only have a perturbative solution in the matching expansion.  As we saw
in the previous section, perturbative solutions have apparent CTCs and
are not suitable for such analysis.

To work around this problem, we will instead make use of supertube physics to
argue that all the parameters are constrained and thus our non-Abelian
solution represents a bound state. 
Actually, we can fix all the parameters from this argument.
It is not a rigorous argument, but
is robust enough to give convincing evidence that the solution represents
a bound state.

\subsection{An  argument for a  bound state}
\label{ss:arg_bnd_state}

We know that the monodromy matrices of the two supertubes sitting at
$z=\pm L$ are
\begin{align}
 M_{L}=\begin{pmatrix}1&0\\ -2&1\end{pmatrix},\qquad
 M_{-L}=\begin{pmatrix}3&2\\ -2&-1\end{pmatrix}.
\end{align}
In Appendix \ref{ss:puffed-up_dip_chg_1/4-BPS_ctr}, we derived the
monodromy matrix of the supertube produced by the supertube transition of
a general 1/4-BPS codimension-3 center.  In the one-modulus class that
we are working in ($\tau^1=\tau^2=i$, $\tau^3$: any), a general
1/4-BPS codimension-3 center has charge $\Gamma={g_s l_s\over
2}(a,(b,b,c),(d,d,a),-{c\over 2})$, where $a,b,c,d\in\bbZ$, $ad+bc=0$
and not all of $a,b,c,d$ simultaneously vanish.  Using the formulas
\eqref{monodromy_ac} and \eqref{monodromy_bd}, it is easy to see that
the unique sets of charges that lead to supertubes with monodromy $M_{\pm
L}$ are the ones with
\begin{align}
 M_L: &~ c=d=0,\qquad\qquad
 M_{-L}: ~ a=-c,~b=d,\label{cond_abcd_2-tube}
\end{align}
with the dipole charge $q=2$ for both cases.
In terms of complex charges (cf.~\eqref{QFQG_q'n}),
\begin{align}
 Q_F&={g_s l_s \over 2}(c+id) ,
 \qquad Q_G={g_s l_s \over 2}(a-ib),
\end{align}
the condition \eqref{cond_abcd_2-tube} can be written as:
\begin{align}
 M_L   :~~ Q_F=0,\qquad\qquad
 M_{-L}:~~  Q_F=-Q_G.
\label{QFQGcond_2tubes}
\end{align}
The supertubes at $z=\pm L$ must have come from two codimension-3
centers with charges satisfying this condition,
respectively.\footnote{To be precise, by charges here, we mean Page
charges discussed in Section \ref{ss:geom_chg}.}

From \eqref{QFQG_def}, the total charges of our two-supertube
configuration is
\begin{align}
 \begin{pmatrix}Q_F\\ Q_G \end{pmatrix}_{\!\text{total}}
 =c\sqrt{R}
 \begin{pmatrix} i\nu \\ 1 \end{pmatrix}.
\end{align}
Let us split this total charge into the ones for the
$z=+L$ supertube and the ones for the $z=-L$ supertube as
\begin{align}
 \begin{pmatrix}Q_F\\ Q_G \end{pmatrix}_{\!\text{total}}
 =
 \begin{pmatrix}Q_F\\ Q_G \end{pmatrix}_L
 +
 \begin{pmatrix}Q_F\\ Q_G \end{pmatrix}_{-L},
\end{align}
and require that the individual charges satisfy the condition
\eqref{QFQGcond_2tubes}, namely,
\begin{align}
 Q_{F,L}=0,\qquad\qquad Q_{F,-L}=-Q_{G,-L}.
\end{align}
We immediately find
\begin{subequations}
\label{QFQG_indiv}
 \begin{align}
 \begin{pmatrix}Q_F\\ Q_G \end{pmatrix}_{L}~
 &=
 c\sqrt{R} \begin{pmatrix}0\\ 1+i\nu \end{pmatrix}
\label{QFQG_indiv1}
 ,\\
 \begin{pmatrix}Q_F\\ Q_G \end{pmatrix}_{-L}
 &=
 c\sqrt{R} \begin{pmatrix}i\nu \\ -i\nu \end{pmatrix}.
\label{QFQG_indiv2}
 \end{align}
\end{subequations}
In our solution we have two codimension-2 supertubes, instead of codimension-3
centers.  However, these supertubes must still carry the original
monopole charges \eqref{QFQG_indiv} dissolved into their worldvolume.
Using the relation \eqref{tau12harm}, we can express
\eqref{QFQG_indiv} in terms of charges vectors as
\begin{align}
 \Gamma_{\pm L}
 =\biggl(\Re Q_G,(-\Im Q_G,-\Im Q_G,\Re Q_F),(\Im Q_F,\Im Q_F,\Re Q_G),-\half\Re Q_F\biggr)_{\pm L}.
 \label{GammapmL}
\end{align}

The radii and angular momentum of the configuration are determined by the charges of the centers. Then, we can study what the radii of the circular supertubes generated by the supertube
transition of codimension-3 centers with charges \eqref{QFQG_indiv} are.
This has been worked out in Appendix \ref{app:R,J_supertube} and, using
the formula \eqref{eq:R,J_supertube}, it is not difficult to show that
the radii of the supertubes at $z=\pm L$ are given by
\begin{align}
\begin{split}
 \cR_L^2&=R|c(1+i\nu)|^2
 =R|c|^2\biggl[1+{2l\over \pi}+{1\over \pi^2}\biggl(\biggl(\log{4R\over |L|}\biggr)^2+l^2\biggr)\biggr]
 \,,\\
 \cR_{-L}^2&=R|c|^2 |\nu|^2
  ={R|c|^2\over \pi^2}\biggl(\biggl(\log{4R\over |L|}\biggr)^2+l^2\biggr)
 \,.
\end{split}
\label{R_isolation}
\end{align}
In deriving this, each supertube was assumed to be in isolation; the
actual radii must be corrected by the interaction between the two tubes.
On the other hand, the radii squared of the two tubes in our actual
solution are
\begin{align}
 (R\pm \Re L)^2=(R\pm |L|\cos l)^2.
 \label{R_actual}
\end{align}
As a preliminary, zeroth-order approximation, let us equate
\eqref{R_isolation} and \eqref{R_actual}. It is not difficult to show
that, unless $l=-{\pi\over 2}$, there is no solution that is consistent with the
colliding limit, ${R\over |L|}\gg 1$.  If $l=-{\pi\over 2}$, the two
supertubes have the same radius and the condition that 
\eqref{R_isolation} equals \eqref{R_actual} gives
\begin{align}
   |c|
 =\frac{\sqrt{R}}{\abs{\nu}}={\pi\sqrt{R}\over \sqrt{\big(\log{4R\over |L|}\big)^2+{\pi^2\over 4}}}.
\end{align}
The total charges \eqref{QFQG_def} are, if we set $c=|c| e^{i\gamma}$,
\begin{align}
 (Q_F,Q_G)=c\sqrt{R}\,(i\nu,1)
 ={e^{i\gamma}R\over \sqrt{\big(\log{4R\over |L|}\big)^2+{\pi^2\over 4}}}
 \left(i\log{4R\over |L|}-{\pi\over 2},\pi\right).
\end{align}
Fixing these charges will fix $\gamma,R,|L|$.  So, everything is fixed.

In summary, consideration of supertube physics suggests that the
configurational parameters of our two-supertube solution are all fixed
if we fix the asymptotic charges.  In particular, it is impossible to
take the two tubes infinitely far apart.  This is strong evidence that our solution is
a bound state.  Having the same asymptotic charges as a
black hole with a finite horizon, it should represent a microstate of a
genuine black hole.
Our argument is not rigorous in the sense that, in computing the
supertube radii \eqref{R_isolation}, we ignored the interaction between
the tubes.  Therefore, precise values such as $l=-{\pi\over 2}$ may not
be reliable.  However, we expect that it captures the essential physics
and the conclusion remains valid even for more accurate treatments.

\subsection{A cancellation mechanism for angular momentum}

In the last section, we pointed out the puzzling fact that the total
angular momentum of our solution vanishes, even though the two
constituent supertubes are expected to carry non-vanishing angular
momentum.  Here, we argue that this is due to cancellation between the
angular momentum $J_{\pm L}$ carried by the two individual tubes and the
angular momentum $J_{\text{cross}}$ that comes from the electromagnetic
crossing between the two tubes; namely,
\begin{align}
 J_\text{total}=J_{L}+J_{-L}+J_{\text{cross}}\approx 0.
\label{Jtot=0_breakdown}
\end{align}
Just as in Section \ref{ss:arg_bnd_state}, our argument will not be
rigorous; we will see that \eqref{Jtot=0_breakdown} holds only to the
leading order in $|L|\over R$.  We expect that, in an exact treatment,
\eqref{Jtot=0_breakdown} will hold as a precise equality. However, this
study is beyond the scope of this paper.

In our solution, we have two round supertubes which were produced by the
supertube effect of codimension-3 centers with charges
\eqref{QFQG_indiv}.  In Appendix \ref{app:R,J_supertube}, we computed
the angular momentum carried by a round supertube created from a general
1/4-BPS codimension-3 center.  Applying the formula
\eqref{eq:R,J_supertube} to the charges \eqref{QFQG_indiv}, it is not
difficult to show that  the component of angular momentum along the
axis of the tubes ($x^3$-axis) is\footnote{The sign was determined from
the sign of $\omega_2=\omega_\phi/R$ in \eqref{omega_2D} near $z=\pm L$
using \eqref{apaDp_zL} and \eqref{apaDp_zmL}.}
\begin{align}
 J_L&
 =-{R|c|^2 (1+|\nu|^2-2\Im\nu)\over 4G_4},
 \qquad\qquad
 J_{-L}
 =-{R|c|^2 |\nu|^2\over 4G_4}.
\end{align}

Now let us turn to $J_{\text{cross}}$.  For multi-center codimension-3
solutions with charge vectors $\Gamma^p$, there is non-vanishing angular
momentum coming from the crossing  between electric and magnetic fields
given by  \cite{Denef:2000nb}
\begin{align}
 {\bf J}_{\text{cross}}={1\over 2G_4}\sum_{p<q}\bracket{\Gamma^p,\Gamma^q}
 {{\bf a}_{pq}\over |{\bf a}_{pq}|},\qquad
 {\bf a}_{pq}\equiv{\bf a}_p-{\bf a}_q.
\label{JpmL}
\end{align}
In the present case, we have supertubes with codimension 2, not 3.
However, let us still apply this formula using the tubes' monopole
charges \eqref{QFQG_indiv} (or \eqref{GammapmL}) . This is not precise,
but must give a rough approximation of the crossing angular momentum for
our solution.  Using \eqref{QFQG_indiv} and \eqref{GammapmL}, the
component of the angular momentum along the tube axis is\footnote{In
Section \ref{ss:arg_bnd_state}, we argued that the physically allowed
configuration in the limit ${R\over |L|}\gg 1$ has $l=-{\pi\over 2}$,
which means that the center of the $z=\pm L$ tubes are at $x^3=\mp |L|$.  This determines the sign of \eqref{J_cross}.}
\begin{align}
 J_{\rm cross}={1\over 2G_4}\bracket{\Gamma_{-L},\Gamma_{L}}
=-{R|c|^2(\Im \nu -|\nu|^2)\over 2G_4}.
\label{J_cross}
\end{align}

If we add
\eqref{J_cross} and
\eqref{JpmL}, we get
\begin{align}
 J_L+J_{-L}+J_{\rm cross}
 =-{R |c|^2\over 4G_4}.
\label{Jtot_actual}
\end{align}
This is much smaller than  the individual
terms:
\begin{align}
  J_L,J_{-L},J_{\rm cross}
 \sim {R|c|^2|\nu|^2\over G_4}
 \sim {R|c|^2 (\log{R\over |L|})^2\over G_4}
\end{align}
because we are taking the limit ${R\over |L|}\gg 1$.  Therefore, we
conclude that \eqref{Jtot=0_breakdown} holds to the leading order in
$|L|\over R$.

This is an interesting observation, suggesting that the vanishing of
angular momentum in our configuration is indeed due to cancellation
between the ``tube'' angular momentum and the ``cross'' angular
momentum.  Presumably, the nonzero reminder \eqref{Jtot_actual} gets
canceled if we take into account the contribution to the angular
momentum arising from the interaction between the two tubes (recall that we
computed the angular momentum of supertubes as if they were in isolation).

\section{Future directions}

We constructed our solution by taking the configuration that appeared in the $\LG{SU}(2)$ Seiberg-Witten theory as the near-region solution.
More specifically, it was a holomorphic fibration of a genus-1 Riemann
surface on a base of complex dimension 1.  However, this is just an
example, so any other such holomorphic fibration will work.  In
particular, any F-theory solution can be used for the near-region
solution.  In the standard F-theory background, the metric only knows
about the torus modulus $\tau$, but in our case we also need the periods
$(a_D,a)$ and richer structure is expected.
We can generalize this structure by replacing the torus fiber by a
higher-genus Riemann surface.  For example, if one considers
compactification of type IIA on $T^2\times K3$, the U-duality group
becomes $\LG{O}(22,6;\bbZ)$, which contains the genus-2 modular group
$\LG{Sp}(4,\mathbb{R})$. Therefore, one can construct configuration of more general
supertubes using a fibration of a genus-2 Riemann surface over a
base \cite{Braun:2013yla}.
One can also consider generalizing the base.  In the near region the
base is complex 1-dimensional, while in the far region it is real
3-dimensional.  By including an internal $S^1$ direction, one can
extend the base to a complex 2-dimensional space, where a supertube
must appear as a complex curve around which there is a monodromy of the
fiber.  In such a setup, one can use the power of complex analysis and
it might help to construct solutions on a real 3-dimensional base as the
one we encountered in the current paper.

It is known that the geometry of the Seiberg-Witten theory has a string
theory realization \cite{Sen:1996vd, Dasgupta:1996ij, Banks:1996nj}.  If
one realizes the Seiberg-Witten curve as a configuration of F-theory
7-branes, then the worldvolume theory of a probe D3-brane in that
geometry is exactly the $d=4,\cN=2$ theory.  One may wonder if our
solution also represents a moduli space of some gauge theory on a probe
D-brane.  However, such interpretation does not seem straightforward.
The near-region geometry looks very similar to F-theory configurations,
but the 7-brane in the current setup is not just a pure 7-brane but it
has some worldvolume fluxes turned on to carry 5-brane and 1-brane
charges.  Therefore, it is not immediately obvious what probe brane one
should take.  Furthermore, although the near-region configuration
preserves 16 supersymmetries, only 4 supersymmetries are preserved in
the far region, as a 4-charge black-hole microstate. A brane probe will
most likely halve the supersymmetries in each region.  So, the relevant
theory seems to be $d=3,\cN=1$ (or $d=2,\cN=2$) theory whose moduli
space has a special locus, which corresponds to the near region, at
which supersymmetry is enhanced to $\cN=4$ (or $\cN=8$).  It is
interesting to investigate what the theory can be.

We developed techniques to construct solutions in the far and near
regions separately and connect them by a matching expansion.  We worked
out only first terms in the expansion, but one can in principle carry
out this to any order.  In some situations one may be able to carry out
the infinite sum and obtain the exact solution in entire $\bbR^3$.  Such
exact solutions are important because, as discussed below
\eqref{mepq19Aug17}, there are features of the exact solution that are
not visible at any finite order.  Such features include the precise
structure of the monodromy and the metric near the supertubes.  They are
crucial to analyze the no-CTC condition near the supertubes and fix
parameters of the solution, such as $L$ and $R$. We hope to be able to
report development in that direction in near future \cite{nextpaper}.

In this paper, we mainly considered the case where two of the three
moduli are frozen.  It is interesting to investigate possible solutions
in the case where this assumption is relaxed.  In Appendix
\ref{app:tau1=i}, we discussed the case where two moduli are dynamical.
For example, it is interesting to study how the solutions studied in
\cite{Park:2015gka} fit in the formulation developed in Appendix
\ref{app:tau1=i}\@.
Relatedly, we assumed that in the near region the modulus $\tau^3$ is
holomorphic.  However, as far as supersymmetry is concerned, this is not
necessary; the only requirement is that the harmonic functions be
written as a sum of holomorphic and anti-holomorphic functions.  It
would be interesting to see if there are physically allowed solutions
for which $\tau^3$ is not holomorphic.

Our configuration has the same asymptotic charge as a 4D black hole.  4D
black holes are often discussed in the context of the AdS$_3$/CFT$_2$
duality where the boundary CFT is the so-called MSW CFT
\cite{Maldacena:1997de}.  However, this CFT is not as well-understood as
the D1-D5 CFT which appears as the dual of black-hole systems in 5D\@.
It is interesting to see if our solutions can be generalized to
construct a microstate for 5D black holes; for recent work to relate
microstates of the MSW CFT and those of the D1-D5 CFT, see
\cite{Bena:2017geu}.

\section*{Acknowledgments}

\vspace{-2mm}

We thank Iosif Bena, Eric Bergshoeff, Stefano Giusto, Oleg Lunin, Takahiro
Nishinaka, Eoin \'O Colg\'ain, Kazumi Okuyama, Rodolfo Russo, Nicholas
Warner for useful discussions.
This work was supported in part by the Science and Technology Facilities
Council (STFC) Consolidated Grant ST/L000415/1 ``String theory, gauge
theory \& duality'', JSPS KAKENHI Grant Number JP16H03979,  MEXT
KAKENHI Grant Numbers JP17H06357 and JP17H06359, JSPS Postdoctoral Fellowship
and Fundaci\'on S\'eneca/Universidad de Murcia (Programa Saavedra Fajardo).
JJFM and MP are grateful to Queen Mary University of London for
hospitality.
We would like to thank the Yukawa Institute for Theoretical Physics at
Kyoto University for hospitality during the workshop YITP-W-17-08
``Strings and Fields 2017,'' where part of this work was carried out.

\appendix


\section{Duality transformation of harmonic functions}
\label{app:duality_H}

In Section \ref{sec:harmonic}, we showed that the $[\text{SL}(2,\bbZ)]^3$
duality of the STU model acts on harmonic functions as
\eqref{SL2transf_harmfunc}.  Here, we discuss some aspects of the
duality transformation.

\bigskip

In the main text, we introduced vectors such as $H=\{V,K^I,L_I,M\}$.  To
see the group theory structure, it is more convenient to introduce the
$\LG{Sp}(8,\mathbb{R})$ vector \cite{DallAgata:2010srl}\begin{align}
 \cH=(\cH^\Lambda,\cH_\Lambda)
 =(\cH^0,\cH^I,\cH_0,\cH_I)={1\over\sqrt{2}}(-V,-K^I,2M,L_I)
 \label{cH_vs_H}
\end{align}
which transforms in the standard way under the four-dimensional
electromagnetic $\LG{Sp}(8,\mathbb{R})$ duality transformation of $\cN=2$ supergravity.

The skew product $\bracket{H,H'}$ defined in \eqref{def_<H,H>} can be
written as
\begin{align}
 \bracket{H,H'}=-\cH^\Lambda \cH'_\Lambda+\cH_\Lambda
 \cH'^\Lambda
\end{align}
For a generic $\LG{Sp}(8,\mathbb{R})$ symplectic vector $\cV=(\cV^\Lambda,
\cV_\Lambda)=(\cV^0,\cV^I,\cV_0,\cV_I)$, the quartic invariant
$\cJ_4(\cV)$ is given by
\begin{equation}
\cJ_4 (\cV) = -(\cV^\Lambda \cV_\Lambda)^2
+ 4 \sum_{I<J} \cV^I \cV_I \cV^J \cV_J
-4 \cV^0 \cV_1 \cV_2 \cV_3
+4 \cV_0 \cV^1 \cV^2 \cV^3
%
%
.\label{J4_def_app}
\end{equation}
Using this, the quantity $\cQ$ defined in \eqref{Q_def} and rewritten in
\eqref{J4_def0} can be expressed as
\begin{align} 
 \cQ=J_4(H)=\cJ_4(\cH).
 \label{Q=J4(H)}
\end{align}

In this language, the most general U-duality
transformation can be written as
an $8\times 8$ matrix $S\in[\LG{SU}(1,1)]^3\cong[\LG{SL}(2,\mathbb{R})]^3\subset\LG{Sp}(8,\mathbb{R})$
\cite{Behrndt:1996hu, DallAgata:2010srl}
\begin{align}
S = \cS \cT \cU
\, ,
\label{Smatrix}
\end{align}
where
{\small
\begin{subequations} 
\label{STUmatrices}
 \begin{align}
 \cS
 &=\
 \begin{pmatrix}
 \delta_1 & \gamma_1 & & & & & & 
 \\
 \beta_1 & \alpha_1 & & & & & & 
 \\
 &  & \delta_1 & & & & & \gamma_1
 \\
 & &  & \delta_1 &  & & \gamma_1 &
 \\
 & &  &  & \alpha_1  & -\beta_1 &  &
 \\
 & &  &  & -\gamma_1  & \delta_1 &  &
 \\
 & &  & \beta_1 &  & & \alpha_1 &
 \\
 &  & \beta_1 & & & & & \alpha_1
 \end{pmatrix}
 \, ,
 \\
 \cT
 &=\
 \begin{pmatrix}
 \delta_2 & & \gamma_2 & & & & & 
 \\
 &  \delta_2 & & & & & & \gamma_2
 \\
 \beta_2 & & \alpha_2 & & & & & 
 \\
 & &  & \delta_2 &  & \gamma_2 & &
 \\
 & &  &  & \alpha_2 &   & -\beta_2 &
 \\
 & &  & \beta_2 &  & \alpha_2 & &
 \\
 & &  &  & -\gamma_1 &   & \delta_2 & 
 \\
  & \beta_2 & & & & & & \alpha_2
 \end{pmatrix}
 \, ,
 \\
 \cU
 &=\
 \begin{pmatrix}
 \delta_3 & & & \gamma_3 & & & & 
 \\
 &  \delta_3 & & & & & \gamma_3 &
 \\
 & & \delta_3 & & & \gamma_3 & &
 \\
 \beta_3 & & & \alpha_3 & & & & 
 \\
 & &  & & \alpha_3 & &  & -\beta_3
 \\
 & &  \beta_3 &  &&  \alpha_3 & &
 \\
 & \beta_3 &  &  &  &   & \alpha_3 & 
 \\
 &  & & &-\gamma_3 & & & \delta_3
 \end{pmatrix}
 \, .
 \end{align}
\end{subequations}
}%
with $\alpha_I \delta_I-\beta_I \gamma_I=1$, $I=1,2,3$.
It is straightforward to
show that the action of the matrix 
\eqref{Smatrix} on the symplectic vector $(\cH^\Lambda,\cH_\Lambda)$ 
reproduces the transformation law
\eqref{SL2transf_harmfunc}.

\bigskip

The transformation law
\eqref{SL2transf_harmfunc} means that the eight harmonic functions
transform under the ${\bf 2}\otimes {\bf 2}\otimes {\bf 2}$
representation of $[\text{SL}(2,\bbZ)]^3$ as follows:
\begin{align}
\begin{split}
 (\cH^0,\cH^I,\cH_0, \cH_I)
 &={1\over\sqrt{2}}(-V,-K^I,2M,L_I)\\
 &=(
 \cH^{222};\
 \cH^{122},\cH^{212},\cH^{221};\
 - \cH^{111};\
 \cH^{211},\cH^{121},\cH^{112}
 )
\end{split}
\label{2x2x2}
\end{align}
where $\cH^{a b c}$ ($a,b,c=1,2$) transforms as $\cH^{abc}\to
\sum_{a',b',c'}(M_1)^{aa'}(M_2)^{bb'}(M_3)^{cc'}\cH^{a'b'c'}$.
In terms of $\cH^{a b c}$,
\begin{align}
 -\bracket{H,H'}
 &=\cH^\Lambda \cH'_\Lambda - \cH_\Lambda \cH'^\Lambda 
 =\epsilon_{a_1 a_2}\epsilon_{b_1 b_2}\epsilon_{c_1 c_2}
 \cH^{a_1 b_1 c_1}\cH^{a_2 b_2 c_2}
 ,\\
 J_4(H)&=\cJ_4(\cH)
 =\half
 \epsilon_{a_1 a_2}\epsilon_{a_3 a_4}
 \epsilon_{b_1 b_2}\epsilon_{b_3 b_4}
 \epsilon_{c_1 c_3}\epsilon_{c_2 c_4}
 \cH^{a_1 b_1 c_1}\cH^{a_2 b_2 c_2} \cH^{a_3 b_3 c_3}\cH^{a_4 b_4 c_4}.
\end{align}

A matrix $M^{ab}$ cannot be written as a product of two vectors
$u^a,v^b$ in general but it can be written as a sum of multiple
vectors, $M^{ab}=\sum_i u^a_i  v^b_i$.  Similarly, we must be
able to decompose the tensor $\cH^{a b c}$ as
\begin{align}
 \cH^{a b c}=\sum_i u^{a}_i v^{b}_i w^c_i,\label{fufq10Jul17}
\end{align}
where $u^{a}_i$, $v^{b}_i$, and $w^c_i$ are real functions transforming
as doublets of $\text{SL}(2,\bbZ)_1$, $\text{SL}(2,\bbZ)_2$, and $\text{SL}(2,\bbZ)_3$,
respectively.

Let us consider the situation considered in Appendix \ref{app:tau1=i}
where we set one of the moduli to a trivial value: $\tau^1=i$.  Here we
will give an alternative proof that the harmonic functions in this case
are given by \eqref{igra7Jul17}, \eqref{igrd7Jul17}.  As we can see in \eqref{netr6Jul17}, the combinations of harmonic functions
that transform nicely under the remaining $\text{SL}(2,\bbZ)_2\times
\text{SL}(2,\bbZ)_3$ are $V-iK^1$, $K^2+iL_3$, $K^3+iL_2$ and $-L_1-2iM$.  In terms of
$\cH^{a b c}$, they are
\begin{align}
\begin{split}
 V-iK^1   &=\sqrt{2}\,(-\cH^{222}+i\cH^{122})\equiv \cH^{22},\\
 K^2+iL_3 &=\sqrt{2}\,(-\cH^{212}+i\cH^{112})\equiv \cH^{12},\\
 K^3+iL_2 &=\sqrt{2}\,(-\cH^{221}+i\cH^{121})\equiv \cH^{21},\\
 -L_1-2iM &=\sqrt{2}\,(-\cH^{211}+i\cH^{111})\equiv \cH^{11}.
\end{split}
\end{align}
The components of the tensor $\cH^{b c}$ defined here are complex functions transforming as
a ${\bf 2}\otimes {\bf 2}$ of $\text{SL}(2,\bbZ)_2\times \text{SL}(2,\bbZ)_3$.  Just
as in \eqref{fufq10Jul17}, we can decompose it as
\begin{align} 
 \cH^{b c}=\sum_i V^{b}_i W^c_i,\label{fvfc10Jul17}
\end{align}
where $V^b_i,W^c_i$ are complex.  However, 
this is inconsistent with the constraint \eqref{neqx6Jul17}, which reads
in terms of $\cH^{b c}$ as
\begin{align}
 \cH^{11}\cH^{22}=\cH^{12} \cH^{21},
\end{align}
unless the summation over $i$ in \eqref{fvfc10Jul17} has only one term.
In that case,
\begin{align}
\begin{split}
 V-iK^1   &=\cH^{22}=V^2 W^2,\qquad
 K^2+iL_3 =\cH^{12}=V^1 W^2,\\
 K^3+iL_2 &=\cH^{21}=V^2 W^1,\qquad
 -L_1-2iM =\cH^{11}=V^1 W^1.\\ 
\end{split}
\end{align}
This is the same as \eqref{igra7Jul17}, \eqref{igrd7Jul17} with the
identification
$\big(\begin{smallmatrix}V^1\\V^2\end{smallmatrix}\big)=
\big(\begin{smallmatrix}F_2\\G_2\end{smallmatrix}\big)$,
$\big(\begin{smallmatrix}W^1\\W^2\end{smallmatrix}\big)=
\big(\begin{smallmatrix}F_3\\G_3\end{smallmatrix}\big)$.

\bigskip

It is interesting to see how the transformations of the harmonic
functions known in the literature are embedded in the general
$[\text{SL}(2,\bbZ)]^3$ transformation \eqref{SL2transf_harmfunc}.  We will
consider the ``gauge transformation'' \cite{Bena:2005ni} and the
``spectral flow transformation'' \cite{Bena:2008wt} as such
transformations.  To our knowledge, explicit $[\text{SL}(2,\bbZ)]^3$
matrices for these transformations have not been explicitly written down
in the literature.  For a discussion on how these transformations are
embedded in the U-duality group of the STU model from a different
perspective, see \cite{DallAgata:2010srl}.

The so-called ``gauge transformation'' \cite{Bena:2005ni} is defined as
the following transformation of harmonic functions:
\begin{align}
\begin{aligned}
V&\to V , \\
 K^I  &\to K^I + c^I V ,
\\
L_I &\to L_I 
-C_{IJK} c^J K^K
-\frac12 C_{IJK} c^J c^K V ,
\\
M 
&\to
M
- \frac12 c^I L_I
+\frac{1}{4} C_{IJK} c^I c^J  K^K
+\frac{1}{12} C_{IJK} c^I c^J c^K V .
\end{aligned}
\label{eq:gaugetransf}
\end{align}
It is easy to see that this transformation is a special case of general
$[\text{SL}(2,\bbZ)]^3$ transformations \eqref{SL2transf_harmfunc} with
\begin{align}
M_I=\begin{pmatrix}1&c^I \\ 0&1\end{pmatrix},\qquad I=1,2,3.
\end{align}
This transformation shifts the $B$-field as
\begin{align}
B_2
\to 
B_2 
+{c^1 \ap\over R_4 R_5}J_1
+{c^2 \ap\over R_6 R_7}J_2
+{c^3 \ap\over R_8 R_9}J_3
.
\label{gauge_trfm_shift_B}
\end{align}

If one likes, the shift in $B_2$, \eqref{gauge_trfm_shift_B}, can be
always undone by subtracting ${c^1 \ap\over R_4 R_5}J_1 +{c^2 \ap\over
R_6 R_7}J_2 +{c^3 \ap\over R_8 R_9}J_3$ from $B_2$ by hand, because
subtracting from $B_2$ the closed form $J_I$ affects none of the
equations of motion or supersymmetry conditions.  This is relevant
especially in 5D solutions (for which $h^0=0$) because, changing the
asymptotic value of $B_2$ as in \eqref{gauge_trfm_shift_B} would mean to
change the asymptotic value of the Wilson loop along $\psi$ for a 5D
gauge field that descends from the M-theory 3-form $A_{\mu ij}$.  Such a
gauge transformation would not vanish at infinity in 5D and is not
allowed. So, one must always undo the shift \eqref{gauge_trfm_shift_B}
after doing the gauge transformation \eqref{eq:gaugetransf}. After this
procedure, no gauge-invariant fields are changed under the
transformation \eqref{eq:gaugetransf} and it is just re-parametrization
of harmonic functions $\{V,K^I,L_I,M\}$.

The ``spectral flow transformation'' is defined as
\cite{Bena:2008wt} 
\begin{align}
\begin{aligned}
V&\to V+\gamma_I K^I-{1\over 2}C^{IJK}\gamma_I \gamma_J L_K
 +{1\over 3}C^{IJK}\gamma_I\gamma_J\gamma_K M ,\\
K^I&\to K^I-C^{IJK}\gamma_J L_K+C^{IJK}\gamma_J \gamma_K M ,\\
 L_I &\to L_I-2\gamma_I M ,\\
 M&\to M ,
\end{aligned}
\label{eq:spectraltransf}
\end{align}
where $C^{IJK}=C_{IJK}$.  This transformation has been used extensively
to generate new solutions from known ones. It is easy to see that this
transformation is a special case of general $\text{SL}(2,\bbZ)$
transformations with
\begin{align}
M_I=\begin{pmatrix}1&0 \\ \gamma_I&1\end{pmatrix},\qquad I=1,2,3.
\end{align}

\section{Matching to higher order}
\label{app:perturb}

In the main text, we worked out the matching between the far- and
near-region solutions to the leading order.  In this Appendix, we carry
out the matching to higher order.

From the large-$|z|$ expansion of the near-region solution
\eqref{apadpinf}, we find that the
far-region solution must have the following expansion:
\begin{subequations}
\label{FGexpn}
 \begin{align}
 F&=\sqrt{\eta-\cos\sigma}\,\sum_{n=0}^\infty e^{-i\frac{4n+1}{2}\sigma}\left(f_n(\eta)-\frac{\sigma}{\pi}g_n(\eta)\right)\,,\\
 G&=\sqrt{\eta-\cos\sigma}\,\sum_{n=0}^\infty e^{-i\frac{4n+1}{2}\sigma}g_n(\eta)\,.
 \end{align}
\end{subequations}
The Laplace equations for $F$ and $G$ lead to
\begin{align}
\begin{split}
 (1-\eta^2)f_n''-2\eta f_n'+2n(2n+1)f_n&=\frac{i}{\pi}(4n+1)g_n\,,
\\
 (1-\eta^2)g_n''-2\eta g_n'+2n(2n+1)g_n&=0
 \,.
\end{split}
\label{fngneqs}
\end{align}
The equation for $g_n$ is the standard Legendre differential equation
while the one for $f_n$ is an inhomogeneous Legendre differential
equation of resonant type \cite{Backhouse:1986rle}.

The general solution for $g_n(\eta)$ is given by
\begin{align}
g_n(\eta)=A_{2n}P_{2n}(\eta)+B_{2n}Q_{2n}(\eta)\,,
\end{align}
where $P_{2n}(\eta)$ is the Legendre polynomial and $Q_{2n}(\eta)$ is
the Legendre function of the second kind.  As $Q_{2n}(\eta)$ diverges at
3D infinity and on the $x^3$-axis (see Footnote \ref{ftnt:Q_div}), we require $B_{2n}=0$. The expression for
$P_{2n}(\eta)$ for some small values of $n$ is
\begin{subequations}
\begin{align}
P_0(\eta)&=1\,,\\
P_2(\eta)&=\frac{1}{2}(3\eta^2-1)\,,\\
P_4(\eta)&=\frac{1}{8}(35\eta^4-30\eta^2+3)\,.
\end{align}
\end{subequations}
$P_{2n}(\eta)$ are normalized so that $P_{2n}(1)=1$.

Having found $g_n$, we can plug it into \eqref{fngneqs} to find $f_n$.
We have not been able to find a simple explicit expression for $f_n$ that
works for general $n$. We give the following integral form:
\begin{align}
f_n(\eta)&=C_{2n}P_{2n}(\eta)+D_{2n}Q_{2n}(\eta)\notag\\
&\quad-\frac{i}{\pi}A_{2n}(4n+1)
\left(
P_{2n}(\eta)\int_1^\eta ds\, P_{2n}(s)Q_{2n}(s)-Q_{2n}(\eta)\int_1^\eta ds\, [P_{2n}(s)]^2
\right).
\end{align}
We have chosen the particular solution (the last term) to
vanish at 3D infinity ($\eta=1$). As before, we require
$D_{2n}=0$ so that $f_n$ is finite at infinity.  For given $n$, it is
easy to carry out the integral and the explicit expression for a few
small values of $n$ is
\begin{subequations}
\label{first_fn}
\begin{align}
f_0(\eta)&=C_0-\frac{i}{\pi}A_0\ln\frac{\eta+1}{2}\,,\\
f_1(\eta)&=C_2P_2(\eta)-\frac{i}{\pi}A_2\left(P_2(\eta)\ln\frac{\eta+1}{2}+\frac{1}{4}(\eta-1)(7\eta+1)\right)\,,\\
f_2(\eta)&=C_4P_4(\eta)-\frac{i}{\pi}A_4\left(P_4(\eta)\ln\frac{\eta+1}{2}+\frac{1}{96}(\eta-1)(533\eta^3+113\eta^2-241\eta-21)\right)\,.
\end{align}
\end{subequations}

The undetermined coefficients $A_{2n}$ and $C_{2n}$ are fixed by matching the
expansion \eqref{FGexpn} order by order with the large-$|z|$ expansion
of the near-region solution given in \eqref{apadpinf}. This has been
done for the leading $n=0$ term in the main text in Section
\ref{sec:far}; see \eqref{0thcoeff}. For $n=1$, this determines the
coefficients to be
\begin{align}
A_2=\frac{cL^2}{2(2R)^{5/2}}\,,\quad\qquad C_2=\frac{i}{\pi}\frac{cL^2}{2(2R)^{5/2}}\left(\ln\frac{4R}{L}-\frac{1}{2}\right)\,.
\end{align}

%
%
%
%

\section{Configurations with only two moduli}
\label{app:tau1=i}

Let us consider configurations with one modulus set to a trivial value.
Specifically, we set
\begin{align}
 \tau^1=i,\qquad \tau^2,\tau^3:\text{arbitrary}.
\end{align}
This choice
fixes two harmonic functions; from Eq.~\eqref{kahlerm}, we find
\begin{align}
 -L_1-2iM={(K^2+iL_3)(K^3+iL_2)\over V-iK^1}.\label{neqx6Jul17}
\end{align}
Only six harmonic functions are independent.
In this case, the expression for the other moduli $\tau^{2,3}$ simplifies to
\begin{align}
\tau^2
=
\frac{K^2+ i L_3}{V- i K^1}
\, ,\qquad
\tau^3=
\frac{K^3+ i L_2}{V- i K^1}
\, .\label{netr6Jul17}
\end{align}
Because $\tau^2$ undergoes linear fractional transformation under
$\text{SL}(2,\bbZ)_2$, we can set\footnote{Actually, one could more generally
set $K^2+ i L_3=\sum_i H_2^{(i)} F_2^{(i)}$, $V- i K^1=\sum_i H_2^{(i)}
G_2^{(i)}$ where $\Big(\begin{smallmatrix} F_2^{(i)} \\
G_2^{(i)}\end{smallmatrix}\Big)$ transforms as a doublet under
$\text{SL}(2,\bbZ)_2$ for all $i$.  However, $\tau^1$ would not be invariant
under $\text{SL}(2,\bbZ)_3$, unless the $i$ summation contains only one term. 
For a different argument for \eqref{igra7Jul17}, \eqref{igrd7Jul17}, see Appendix \ref{app:duality_H}.\label{ftnt:tau2=tau3=i}}
\begin{align}
K^2+ i L_3=H_2 F_2,\qquad 
V- i K^1=H_2 G_2,\label{nejz6Jul17}
\end{align}
where under $\text{SL}(2,\bbZ)_2$ the pair $\big(\begin{smallmatrix} F_2 \\
G_2\end{smallmatrix}\big)$ transforms as a doublet while $H_2$ is invariant.
The quantities $F_2,G_2,H_2$ are complex.  With this choice
\eqref{nejz6Jul17}, $\tau^2$ is invariant under $\text{SL}(2,\bbZ)_3$ as it
should be.  Similarly, because $\tau^3$ undergoes linear fractional
transformation under $\text{SL}(2,\bbZ)_3$, we can set
\begin{align}
K^3+ i L_2=H_3 F_3,\qquad 
 V- i K^1=H_3 G_3,\label{nekd6Jul17}
\end{align}
where under $\text{SL}(2,\bbZ)_3$ the pair $\big(\begin{smallmatrix} F_3 \\
G_3\end{smallmatrix}\big)$ transforms as a doublet while $H_3$ is invariant.
$F_3,G_3,H_3$ are complex.  
Combining \eqref{nejz6Jul17} and \eqref{nekd6Jul17}, we find that
$H_2=G_3$ and $H_3=G_2$ and therefore
\begin{align}
 K^2+iL_3=F_2 G_3,\qquad 
 V-iK^1 = G_2 G_3,\qquad
 K^3+iL_2= G_2 F_3,\label{igra7Jul17}
\end{align}
with which \eqref{neqx6Jul17} becomes
\begin{align}
 -L_1-2iM=F_2 F_3.\label{igrd7Jul17}
\end{align}
The moduli \eqref{netr6Jul17} can now be written as
\begin{align}
 \tau^2={F_2\over G_2},\qquad \tau^3={F_3\over G_3}.
\end{align}

In terms of $F_{2,3},G_{2,3}$, the harmonic functions are
\begin{align}
\begin{aligned}
V &=\Re G_2 G_3\, ,&
K^1 &=-\Im G_2 G_3\, ,&
K^2&=\Re F_2 G_3\, ,&
K^3&=\Re G_2 F_3\, ,\\
L_1 &=-\Re F_2 F_3\, ,&
L_2 &=\Im G_2 F_3\, ,&
L_3&=\Im F_2 G_3\, ,&
M&=-\frac{1}{2}\Im F_2 F_3\, .
\end{aligned}
\label{tau3harm}
\end{align}
Because we are parametrizing 6 real harmonic functions using 4 complex
functions $F_{2,3},G_{2,3}$, there is redundancy: the transformation $
\big(\begin{smallmatrix} F_2\\ G_2 \end{smallmatrix}\big) \to H
\big(\begin{smallmatrix} F_2\\ G_2 \end{smallmatrix}\big)$, $
\big(\begin{smallmatrix} F_3\\ G_3 \end{smallmatrix}\big)\to H^{-1}
\big(\begin{smallmatrix} F_3\\ G_3 \end{smallmatrix}\big)$, where $H$ is a
complex function, leaves the harmonic functions invariant. 

Let us consider the no-CTC conditions \eqref{condCTCs}. The condition
\eqref{QcondCTCs} is automatically satisfied because $\cQ = {(K^1 K^3+
L_2 V)^2 (K^1 K^2+L_3 V)^2}/{(({K^1})^2+V^2)^2}\ge0$. The conditions $V
Z_I\ge0$, \eqref{VZcondCTCs}, become
\begin{align}
\begin{split}
 VZ_2&= K^1 K^3+L_2 V =|G_2|^2\Im(F_3\bar{G}_3)=|G_2G_3|^2\Im\tau^3\ge 0
 \, ,
 \\
 VZ_3&= K^1 K^2+L_3 V =|G_3|^2\Im(F_2\bar{G}_2)=|G_2G_3|^2\Im\tau^2\ge 0
 \, .
\end{split}
\end{align}

\section{Supertubes in the one-modulus class}
\label{app:supertubes}

In Section \ref{sec:taucases}, we discussed a class of harmonic
solutions for which only one modulus, $\tau^3=\tau$, is turned on.
(This class is
nothing but a type IIA realization of the solution called the SWIP
solution in the literature \cite{Bergshoeff:1996gg}.)  Here
let us study some properties of supertubes described in this class.

\subsection{Condition for a 1/4-BPS codimension-3 center}
\label{sss:cond_1/4-BPS_codim3_ctr}

Let us consider a codimension-3 center in the harmonic solution and let
the charge vector of the center be $\Gamma$. In terms of quantized
charges, $\Gamma$ can be written as
\begin{align}
 \Gamma={g_s l_s \over 2}\left(a,(b,b,c),(d,d,a),-{c\over 2}\right),
\label{mfak19May17}
\end{align}
where $a,b,c,d\in\bbZ$.  Here, we took into account the constraint
\eqref{tau12iharmcond} and charge quantization~\eqref{charge_q'n}.
In general, this center represents a 1/8-BPS center preserving
4 supercharges, with entropy (see \eqref{S_QFQG})
\begin{align}
 S=2\pi \sqrt{j_4(\Gamma)},\qquad j_4(\Gamma)\equiv (ad+bc)^2.
\end{align}

We would like to find the condition for the charge vector
$\Gamma$ to represent a 1/4-BPS center preserving 8 supercharges, which
can undergo a supertube transition into a codimension-2 center.
According to \cite{Ferrara:1997ci}, a center with charge vector $\Gamma$
represents
\begin{align}
\begin{split}
  \text{4-charge 1/8-BPS center}&~\Leftrightarrow~j_4(\Gamma)>0.\\
 \text{3-charge 1/8-BPS center}&~\Leftrightarrow~
 j_4(\Gamma)=0,~ {\p j_4\over \p x_i}\neq 0\\
 \text{2-charge 1/4-BPS center}&~\Leftrightarrow~
 j_4(\Gamma)={\p j_4\over \p x_i}= 0,~ {\p^2 j_4\over \p x_i\p x_j} \neq 0\\
 \text{1-charge 1/2-BPS center}&~\Leftrightarrow~
 j_4(\Gamma)={\p j_4\over \p x_i}= {\p^2 j_4\over \p x_i\p x_j}=0, ~{\p^3 j_4\over \p x_i\p x_j\p x_k}\neq 0,
\end{split}
\end{align}
where $x_i$ represents charges of D-branes which, in the present case,
are $a,b,c,d$.  Applying this to the present case, we find that
\begin{subequations} 
 \begin{align}
  \text{4-charge 1/8-BPS center}&~~~\Leftrightarrow~~~ ad+bc \neq 0,
 \\
 \text{2-charge 1/4-BPS center}&~~~\Leftrightarrow~~~ ad+bc =0,~~
 \text{but not $a=b=c=d=0$}
 \label{fgpi24Jul17}
 \end{align}
\end{subequations}
In the present class of configurations satisfying \eqref{mfak19May17},
we cannot have a 3-charge 1/8-BPS center or a 1-charge 1/2-BPS center.
For the latter, for example, even if $a=b=c=0$ and $d\neq 0$, it still
represents a D2(45)-D2(67) system which is a 2-charge 1/4-BPS system.

\subsection{Puffed-up dipole charge  for general 1/4-BPS codimension-3 center}
\label{ss:puffed-up_dip_chg_1/4-BPS_ctr}

If the 1/4-BPS system with charges satisfying \eqref{fgpi24Jul17}
polarizes into a supertube, what is its dipole charge, or more
precisely, the monodromy matrix around it?  From \eqref{SL2doublets}, we see that
the combinations of charges that transform as doublets are
\begin{align}
 \begin{pmatrix} K^3 \\ V \end{pmatrix}=
 \begin{pmatrix} -2M \\ L_3 \end{pmatrix}
 \propto
 \begin{pmatrix} c \\ a \end{pmatrix}
 ,\qquad
 \begin{pmatrix} -L_1 \\ K^2 \end{pmatrix}=
 \begin{pmatrix} -L_2 \\ K^1 \end{pmatrix}
 \propto
 \begin{pmatrix} -d \\ b \end{pmatrix}\label{fiei24Jul17}
\end{align}
with $ad+bc=0$.
If we act with a general
$SL(2,\bbZ)$ matrix, the first doublet transforms as
\begin{align}
 \begin{pmatrix} c \\ a \end{pmatrix}
 \to
 \begin{pmatrix} c' \\ a' \end{pmatrix}
 =
 \begin{pmatrix} \alpha & \beta \\ \gamma & \delta \end{pmatrix}
 \begin{pmatrix} c \\ a \end{pmatrix}
 =
 \begin{pmatrix} \alpha c+ \beta a\\ \gamma c +\delta a\end{pmatrix},
\end{align}
where $\alpha,\beta,\gamma,\delta\in\bbZ$ and $\alpha
\delta-\beta\gamma=1$. The second one transforms in the same way. Let us
require that the lower component of the first doublet in
\eqref{fiei24Jul17} vanishes in the transformed frame, namely,
$a'=\gamma c+\delta a=0$.  If we write
\begin{align}
 a=x\hat{a},\qquad c=x\hat{c}, \qquad x=\gcd(a,c),\label{ghgi24Jul17}
\end{align}
so that $\hat{a}$ and $\hat{c}$ are relatively prime, then it is clear
that $a'=0$ for the following choice:
\begin{align}
 \gamma=\hat{a},\qquad \delta=-\hat{c}.
\end{align}
Note that the lower component of the second doublet in
\eqref{fiei24Jul17} also vanishes in the transformed frame:
\begin{align}
 b'=-\gamma d+\delta b=-\hat{a}d-\hat{c}b=-{1\over x}(ad+bc)=0
\end{align}
by the assumption of 1/4-BPSness, \eqref{fgpi24Jul17}.  
For the matrix $\big(\begin{smallmatrix} \alpha&\beta\\
\gamma&\delta\end{smallmatrix}\big)$ to be an $\LG{SL}(2,\bbZ)$ matrix, we must
satisfy
\begin{align}
 \alpha\delta - \beta\gamma=-\alpha \hat{c}-\beta \hat{a}=1, \label{fvcm24Jul17}
\end{align}
but there always exist $\alpha,\beta\in\bbZ$ satisfying this, for
$\hat{a},\hat{c}$ are coprime.

In the frame dualized by the $\LG{SL}(2,\bbZ)_3$ matrix
\begin{align}
 U=
 \begin{pmatrix} \alpha&\beta \\ \hat{a} & -\hat{c} \end{pmatrix}
\end{align}
satisfying \eqref{fvcm24Jul17},
it is easy to show that the charges are
\begin{align}
 \begin{pmatrix} K^3 \\ V \end{pmatrix}=
 \begin{pmatrix} -2M \\ L_3 \end{pmatrix}
 \propto
 \begin{pmatrix} x \\ 0 \end{pmatrix}
 ,\qquad
 \begin{pmatrix} -L_1 \\ K^2 \end{pmatrix}=
 \begin{pmatrix} -L_2 \\ K^1 \end{pmatrix}
 \propto
 \begin{pmatrix} y \\ 0 \end{pmatrix}.
\label{upx7Sep17}
\end{align}
To derive this, we used the fact that, if we write $b,d$ as
\begin{align}
 b=y\hat{b},\qquad d=y\hat{d},\qquad y=\gcd(b,d),
\end{align}
then
the condition $ad+bc=0$
implies that
\begin{align}
 (\hat{b},\hat{d})=
 \pm (\hat{a},-\hat{c}).
\label{wrf7Sep17}
\end{align}

\eqref{upx7Sep17} correspond to the following charges:
\begin{align}
 \text{$x$ units of D4(4567)+D0},\qquad
 \text{$y$ units of D2(45)+D2(67)}.\label{flsq24Jul17}
\end{align}
As we can see from \eqref{codim2supertubes}, both of these pairs must
puff out into ns5($\lambda$4567), where $\lambda$ parametrizes a closed
curve in transverse directions. The $\LG{SL}(2,\bbZ)_3$ monodromy matrix for
ns5($\lambda$4567) is
\begin{align}
 M_{\text{ns5}(\lambda 4567) }=
 \begin{pmatrix}
  1&q\\ 0&1
 \end{pmatrix}
\end{align}
where $q\in\bbZ$ is the dipole charge number (the number of NS5-branes).
If we dualize this back, the monodromy of the supertube in the original
frame is
\begin{align}
 M=U^{-1} M_{\text{ns5}(\lambda 4567) } U
=
 \begin{pmatrix}
  1-q \hat{a}\hat{c}&  q\hat{c}^2 \\
  -q\hat{a}^2 & 1+q\hat{a}\hat{c}
 \end{pmatrix}\label{monodromy_ac}
\end{align}
where we used \eqref{fvcm24Jul17}.
This result is symmetric under the exchange of $c\choose a$ and
$-d\choose b$ as it should be because, 
using \eqref{wrf7Sep17}, we can write this as
\begin{align}
 M=\begin{pmatrix}
    1+q\hat{b}\hat{d} & q\hat{d}^2\\
    -q\hat{b}^2       & 1-q\hat{b}\hat{d}
   \end{pmatrix}.\label{monodromy_bd}
\end{align}

Even in cases where some of $a,b,c,d$ vanish, we can use the formulas
\eqref{monodromy_ac} or \eqref{monodromy_bd}.  If $a=c=0$, we can use
\eqref{monodromy_bd}.  If $b=d=0$, we can use \eqref{monodromy_ac}.  If
$a$ or $c$ vanishes, we can use the rule $\gcd(k,0)=k$ for
$k\in\bbZ_{\neq 0}$ in \eqref{ghgi24Jul17}.  For example, if $c=0$, then
$x=a$ and $\hat{a}=1,\hat{c}=0$.

\subsection{Round supertube}
\label{app:R,J_supertube}

Let us compute the radius and the angular momentum of the round supertube
that is created from a 1/4-BPS center with general $a,b,c,d$ satisfying $ad+bc=0$.

If we T-dualize \eqref{flsq24Jul17} along 7, S-dualize, 
T-dualize along 4567, and then finally S-dualize, we obtain
\begin{align}
 \text{$x$ units of F1(7)+P(7)},\qquad
 \text{$y$ units of F1(6)+P(6)}.\label{jgdm27Jul17}
\end{align}
This is the so-called FP system which is well-studied, rotated in the 67
plane.  In the FP system with F1(7) and P(7) with quantized charges
$N_\text{F1},N_\text{P}\in\bbZ$, the radius $\cR$ and angular momentum $J$ of a
circular configuration are given by (see, e.g.,~\cite{Emparan:2001ux}):
\begin{align}
\cR=l_s{\sqrt{N_\text{F1}N_\text{P}}\over q},\qquad  J={N_\text{F1}N_\text{P}\over q},
\end{align}
where $q\in\bbZ$ is the dipole charge number.  For the rotated system
\eqref{jgdm27Jul17}, this becomes
\begin{align}
\cR=l_s{\sqrt{x^2+y^2}\over q},\qquad J={x^2+y^2\over q}.
\label{eq:R,J_supertube}
\end{align}
Following the duality chain back, we find this expression is again valid for the original frame with general $a,b,c,d\in\bbZ$, $ad+bc=0$.


\section{Harmonic functions for the \boldmath{$\text{D2}+\text{D6}\to 5^2_2$} supertube}
\label{app:D2+D6->522}

In the main text, we reviewed the harmonic functions for the
D2+D2$\to$ns5 supertube \eqref{D2+D2->ns5}.  Here we recall the
harmonic functions for the D2(89)+D6(456789)$\to5_2^2(\lambda$4567;89)
supertube \cite{Park:2015gka}, which is the last line of
\eqref{codim2supertubes}.  This involves the exotic brane $5^2_2$ with a
non-geometric monodromy.

Harmonic functions which describe this supertube are \cite{Park:2015gka}
\begin{align}
\begin{aligned}
V&=f_2\,,\quad 
 &K^1&=\gamma\,,& K^2&=\gamma\,,& K^3&=0\,,\\
&&L_1&=1\,,     & L_2&=1\,,     & L_3&=f_1\,,&\quad
M&=0\,.
\end{aligned}
\end{align}
where $f_1,f_2$ are the same functions that appeared in \eqref{f1,f2_def}.
$\gamma$ is defined in 
\eqref{alpha_gamma} and has the monodromy \eqref{gamma_monodromy}.

The behavior of $V,L_3$ shows that we do have D6(456789) and D2(89)
charges distributed along the profile. On the other hand, the monodromy
can be read off from
\begin{align}
 \begin{pmatrix} -L_1 \\ K^2\end{pmatrix}
 =
 \begin{pmatrix} -1 \\ \gamma\end{pmatrix}
 \to
 \begin{pmatrix} -1 \\ \gamma+1\end{pmatrix}
 =
 \begin{pmatrix} 1 & 0 \\ -1 & 1\end{pmatrix}
 \begin{pmatrix} -L_1 \\ K^2\end{pmatrix}.
\end{align}
From \eqref{SL2transf_harmfunc}, \eqref{SL2doublets}, this means that we
have the following $\text{SL}(2,\bbZ)_3$ monodromy:
\begin{align}
 M_3=\begin{pmatrix}  1 & 0\\ -1 & 1 \end{pmatrix}\in \text{SL}(2,\bbZ)_3.
\end{align}
One can also see this from the K\"ahler moduli,
\begin{align}
\tau^1&=i\sqrt{\frac{f_1}{f_2}}\, ,\qquad
\tau^2=i\sqrt{\frac{f_2}{f_1}}\, ,\qquad
\tau^3=-{1\over \tau'^3},\qquad
\tau'^3=\gamma+i\sqrt{f_1f_2}.\label{tauI_D2D6522}
\end{align}
We see that, as we go once around the supertube, $\tau^{1,2}$ are
single-valued whereas $\tau^3$ has the monodromy
\begin{align}
\tau^3 \to {\tau^3\over -\tau^3+1}.
\end{align}
Because $\tau^3=B_{89}+i\sqrt{\det G_{ab}}$ where $a,b=8,9$, this
monodromy implies that, every time one goes through the supertube, the
radii of the torus $T^2_{89}$ keeps changing.  Namely, this spacetime is
twisted by T-duality and is non-geometric.  This is precisely the
correct monodromy for the $5^2_2$-brane \cite{deBoer:2010ud,
deBoer:2012ma}.

As in the case of the $\text{D2}+\text{D2}\to\mathrm{ns5}$ supertube discussed
around\eqref{D2+D2->ns5}, if $|\dot{\bf F}|=1$, we have $f_1=f_2\equiv f$ and
therefore $\tau^1=\tau^2=i$ as we can see from \eqref{tauI_D2D6522}.
So, the situation reduces to the one-modulus class of Section
\ref{sec:taucases}, with the complex harmonic functions
\begin{align}
 F=i,\qquad G=-i(\gamma+if).\label{FG_D2D6522}
\end{align}


\bibliographystyle{JHEP_MP}
\bibliography{nonabelian_supertubes_paper_refs}


\end{document}